\documentclass[final,3p,times]{elsarticle}

\usepackage{lineno}
\modulolinenumbers[5]

\journal{Journal of .....}

\usepackage{amsmath,amssymb,bm,mathtools}
\usepackage{braket}
\usepackage{siunitx}

\usepackage{graphicx}
\usepackage{subcaption}
\usepackage{booktabs}
\usepackage{multirow}
\usepackage{array}
\usepackage{caption}

\usepackage{algorithm}
\usepackage{algpseudocode}

\usepackage{xcolor}
\usepackage{comment}
\usepackage{cancel}
\usepackage{placeins}
\usepackage{stfloats}
\usepackage{csvsimple}
\usepackage{tcolorbox}
\usepackage{breqn}

\usepackage{hyperref}
\usepackage[capitalise]{cleveref}
\usepackage{xr}

\biboptions{numbers,sort&compress}

\newcommand{\eref}[1]{Eq.~(\ref{#1})}
\newcommand{\fref}[1]{Fig.~\ref{#1}}

\makeatletter
\newcommand*{\addFileDependency}[1]{
  \typeout{(#1)}
  \@addtofilelist{#1}
  \IfFileExists{#1}{}{\typeout{No file #1.}}
}
\makeatother

\begin{document}

\begin{frontmatter}

\title{Bayesian Sequential Quantum Amplitude Estimation for Rare-Event Structural Failure Probability}

\author[1,b]{Alireza Tabarraei\corref{mycorrespondingauthor}}
\cortext[mycorrespondingauthor]{Corresponding author}
\ead{atabarra@charlotte.edu}

\address[1]{Department of Mechanical Engineering and Engineering Science, The University of North Carolina at Charlotte, Charlotte, NC 28223, USA}
\address[b]{School of Data Science, The University of North Carolina at Charlotte, Charlotte, NC 28223, USA}

\begin{abstract}
Structural reliability analysis often requires estimating small failure probabilities under uncertainty, a task for which direct Monte Carlo simulation becomes inefficient because failure observations are scarce. Quantum amplitude estimation offers a potential quadratic improvement in query complexity for bounded expectation estimation, but practical iterative formulations require reliable inference from finite, amplified measurement data. This paper develops a Bayesian sequential formulation of iterative quantum amplitude estimation for rare-event structural failure probability estimation. Structural failure is represented as a binary indicator over a finite stochastic ensemble and encoded through a lookup-table oracle, allowing the failure probability to be treated as a quantum amplitude. Measurement outcomes collected at different Grover depths are assimilated through Bayesian updating over the amplitude angle, yielding posterior estimates, credible intervals, and uncertainty-aware convergence diagnostics. The framework is evaluated on stochastic finite-element benchmark problems, including a one-dimensional bar and an L-bracket with stress concentration. The results show that amplitude amplification converts rare failure events into measurable success probabilities, enabling substantially lower estimation errors than direct Monte Carlo simulation under the same idealized oracle-query budget. The Bayesian formulation achieves point-estimation accuracy comparable to maximum-likelihood IQAE while additionally providing posterior uncertainty quantification, credible intervals, and transparent convergence assessment. The study demonstrates Bayesian IQAE as a statistically interpretable proof-of-concept for quantum-assisted rare-event reliability analysis, while relying on idealized oracle access. 

\end{abstract}


\end{frontmatter}

\section{Introduction}

Reliability analysis has become an essential component of modern computational mechanics, providing a quantitative framework for assessing the effects of uncertainty on structural performance and safety. Engineering systems are routinely subjected to uncertainties arising from material variability, manufacturing defects, geometric imperfections, loading conditions, environmental influences, and modeling assumptions. These uncertainties propagate through computational models and can significantly affect structural response, durability, and failure risk. Consequently, the estimation of failure probabilities has become a central task in the design, certification, and life-cycle management of engineering systems \cite{Madsen1986,Ditlevsen1996,Melchers2018}.

The increasing complexity of computational models has made reliability analysis both more important and more challenging. High-fidelity finite-element simulations, multiscale models, stochastic finite-element formulations, and physics-based digital twins provide unprecedented predictive capabilities, but often at substantial computational cost. As a result, uncertainty quantification and reliability assessment have emerged as active areas of research within computational mechanics \cite{Schueller2007,Ghanem1991,Babuska2004,Xiu2002}. A recurring challenge in these applications is the estimation of rare-event probabilities associated with structural failure. Although such events occur infrequently, they often dominate engineering risk and therefore play a decisive role in safety-critical decision making.

Classical Monte Carlo simulation remains one of the most widely used tools for reliability analysis because of its generality and minimal assumptions regarding the underlying model \cite{Lee2025SubsetSim,Melchers2002,Rubinstein2017,Ang2007,Robert2004}. However, its efficiency deteriorates rapidly as the event of interest becomes rarer. In many practical applications, probabilities of failure may be several orders of magnitude smaller than unity, requiring an extremely large number of model evaluations to achieve acceptable statistical accuracy. This difficulty has motivated the development of numerous advanced reliability methods over the past several decades. Approximation-based approaches such as the First- and Second-Order Reliability Methods (FORM and SORM) provide efficient estimates through local approximations of the limit-state surface \cite{Hasofer1974,Rackwitz1978}. Variance-reduction strategies, including importance sampling, line sampling, subset simulation, and related rare-event simulation techniques, have significantly expanded the range of problems that can be addressed using stochastic simulation \cite{Au2001,Papaioannou2015,Rubinstein2017, Dasgupta2024REIN, Li2025RelaxedSubset}. More recently, surrogate-based approaches employing polynomial chaos expansions, Gaussian-process models, active learning, and Bayesian reliability methods have demonstrated substantial reductions in computational cost for uncertainty propagation and reliability assessment \cite{Sudret2008,Echard2011,Bect2012,Bichon2008,Moustapha2024, kim2024adaptive,
pires2025reliability, Wang2024Kriging}.

Despite these advances, the accurate estimation of extremely small failure probabilities remains computationally demanding \cite{Au2001, Echard2011, Guo2024AdaptiveSampling}. The challenge becomes particularly severe when each model evaluation requires the solution of a large-scale finite-element problem or a computationally intensive multiphysics simulation. Consequently, there is continuing interest in alternative computational paradigms that can improve the efficiency of reliability estimation while maintaining rigorous uncertainty quantification.
Recent progress in quantum computing has stimulated growing interest in quantum algorithms for scientific computing, uncertainty quantification, and risk analysis. Among these developments, quantum amplitude estimation (QAE) has attracted considerable attention because it provides a quadratic improvement in query complexity over classical Monte Carlo sampling for bounded expectation estimation problems \cite{Brassard2002,Montanaro2015}. Since failure probabilities can be represented as expectations of binary failure indicators, amplitude-estimation-based algorithms offer a potentially attractive framework for rare-event reliability analysis. This observation has motivated increasing research activity in quantum risk analysis and uncertainty quantification \cite{Woerner2019}.

The original formulation of quantum amplitude estimation relies on quantum phase estimation and therefore requires deep quantum circuits that are beyond the capabilities of current noisy quantum hardware. To overcome this limitation, several alternative formulations have been developed, including iterative quantum amplitude estimation, maximum-likelihood amplitude estimation, and related likelihood-based approaches that eliminate the need for quantum phase estimation while substantially reducing circuit depth \cite{Suzuki2020,Grinko2021,Nakaji2020,Aaronson2020}. These developments have significantly improved the practical feasibility of amplitude-estimation-based inference and have created new opportunities for applying quantum algorithms to engineering reliability problems.

While existing amplitude-estimation methods have demonstrated promising estimation capabilities, most formulations are designed primarily to estimate unknown probabilities or expectations \cite{Brassard2002, Grinko2021, Suzuki2020, tabarraei2026stabilized}. In reliability engineering, however, uncertainty quantification is often as important as the estimate itself. Engineering decisions are rarely based solely on a predicted probability of failure; they also depend on the confidence that can be placed in that prediction. Confidence intervals, uncertainty bounds, and measures of estimation reliability are fundamental components of risk-informed design and certification procedures.
Bayesian formulations of quantum amplitude estimation have recently attracted growing attention \cite{Koh2020,Ramoa2025,Li2025}. These approaches construct posterior distributions for the underlying quantum amplitude and provide probabilistic alternatives to conventional confidence-interval and maximum-likelihood-based estimators. However, existing developments have been motivated primarily by the amplitude-estimation problem itself and have focused on statistical inference for quantum amplitudes.
The present work adopts a different perspective. Rather than treating the amplitude as the final quantity of interest, Bayesian IQAE is employed as an uncertainty-quantification framework for stochastic structural reliability analysis. The posterior distribution inferred from IQAE measurements is propagated directly to engineering reliability metrics, including failure probabilities and conditional value-at-risk. This enables credible intervals, posterior contraction diagnostics, and uncertainty-aware convergence assessment for quantities that are directly relevant to engineering risk evaluation and decision making.

The present work addresses this need by introducing a Bayesian formulation of iterative quantum amplitude estimation for structural reliability analysis. Rather than representing the unknown failure probability by a single estimate, the proposed framework maintains and sequentially updates a posterior distribution as measurement information becomes available. The resulting formulation naturally provides credible intervals, uncertainty-aware stopping criteria, and a rigorous probabilistic assessment of estimation confidence. In contrast to purely point-estimation approaches, the Bayesian framework explicitly quantifies the uncertainty associated with the inferred failure probability and continuously refines this uncertainty as additional information is acquired.

The methodology is developed within a reliability-analysis framework in which stochastic structural simulations are represented through a binary oracle corresponding to safe and failed realizations. Bayesian updating is performed using measurement outcomes collected at multiple amplification levels, yielding a complete posterior characterization of the underlying failure probability. The resulting framework combines the efficiency of iterative amplitude estimation with the uncertainty quantification capabilities of Bayesian inference.
The numerical investigations assess the performance of Bayesian IQAE for rare-event structural reliability analysis through a series of stochastic finite-element benchmark problems, including explicit quantum-circuit implementations. The study examines estimation accuracy under limited oracle budgets, the evolution of posterior uncertainty as additional computational resources become available, and the influence of event rarity on posterior contraction and confidence measures. The proposed methodology is compared with both classical Monte Carlo simulation and maximum-likelihood iterative quantum amplitude estimation. The results demonstrate that Bayesian IQAE retains competitive estimation accuracy while simultaneously providing a rigorous and interpretable characterization of uncertainty through posterior distributions and credible intervals.


\section{Overview of the Proposed Framework}
\label{sec:framework}

The proposed framework combines stochastic structural simulation, iterative quantum amplitude estimation, and Bayesian inference within a unified reliability-analysis workflow. The objective is to estimate structural failure probabilities and related risk measures while simultaneously quantifying the uncertainty associated with those estimates.

Consider a stochastic structural system characterized by a vector of uncertain parameters $\boldsymbol{\xi}$ with joint probability density function $f_{\boldsymbol{\xi}}(\boldsymbol{\xi})$. For each realization of the uncertain inputs, a computational model is evaluated to obtain a response quantity of interest $Q(\boldsymbol{\xi})$. Failure is defined through the limit-state function
\begin{equation}
g(\boldsymbol{\xi})=
Q_{\rm allow}-Q(\boldsymbol{\xi}),
\label{eq:framework_limit_state}
\end{equation}
where $Q_{\rm allow}$ denotes an allowable response threshold. The corresponding failure probability is
\begin{equation}
p_f=
\mathbb{P}\left[g(\boldsymbol{\xi})<0\right].
\label{eq:framework_pf}
\end{equation}

\begin{figure}[t]
\centering
\includegraphics[width=1.0\linewidth]
{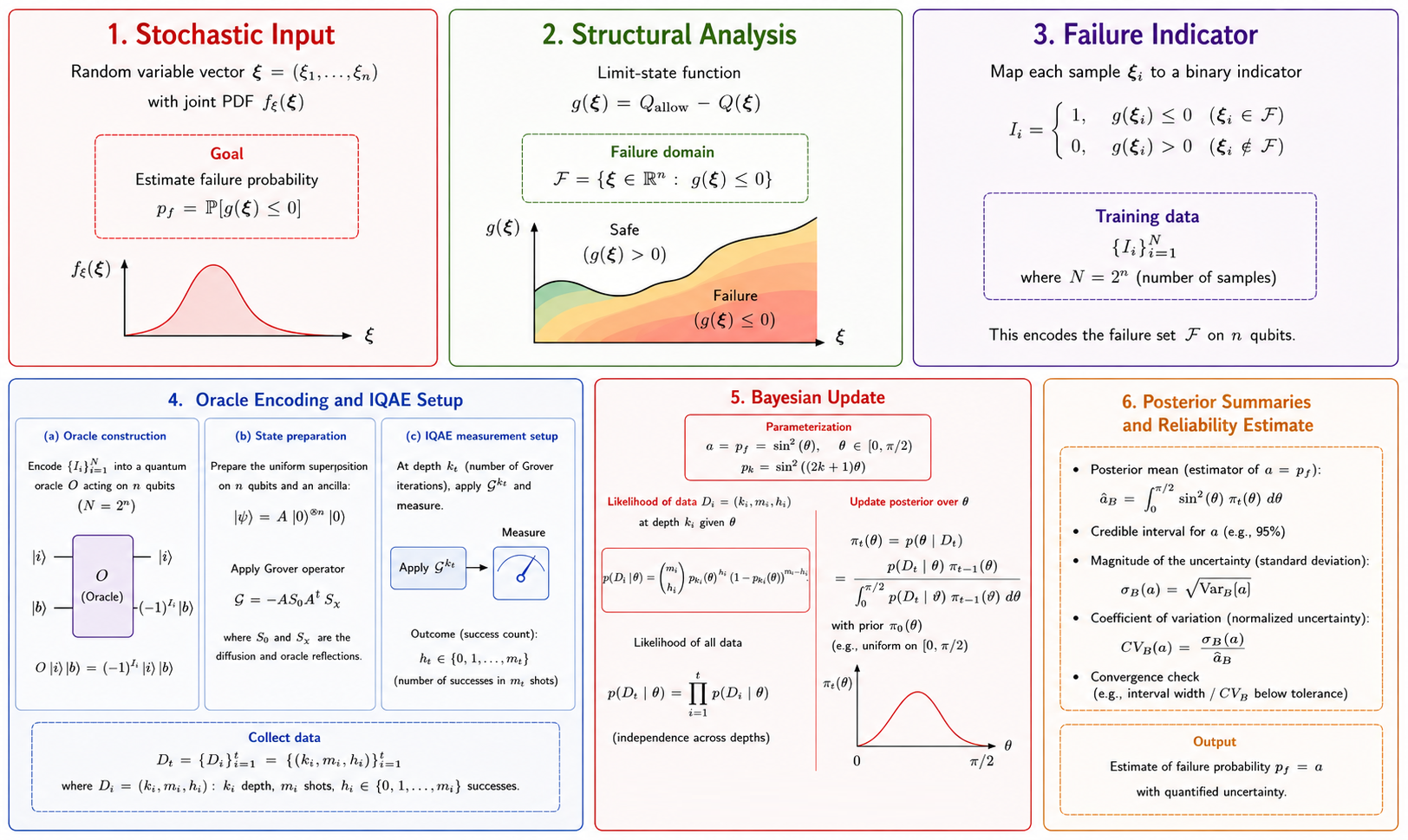}
\caption{
Workflow of the proposed method.
}
\label{fig:workflow}
\end{figure}

The overall workflow is illustrated in Figure~\ref{fig:workflow}. The procedure begins by generating stochastic realizations of the uncertain parameters and evaluating the corresponding structural responses using a classical computational model. The resulting responses are transformed into binary failure indicators that identify whether each realization satisfies the prescribed failure criterion.
The binary failure information is subsequently encoded into a quantum oracle that provides access to the underlying stochastic model. Iterative Quantum Amplitude Estimation (IQAE) is then employed to estimate the unknown failure probability. Measurements are performed at a sequence of amplification depths, producing observation data that contain information about the target amplitude and, consequently, the underlying failure probability.
The measurement outcomes collected throughout the IQAE process are assimilated through Bayesian updating. Starting from an initial prior distribution, the posterior distribution is updated sequentially as additional measurements become available. The resulting posterior provides point estimates of the failure probability together with credible intervals and uncertainty measures that characterize the confidence associated with the estimate.
The final output of the framework consists of a posterior distribution for the reliability quantity of interest, from which posterior means, maximum a posteriori estimates, credible intervals, and convergence diagnostics can be obtained. The same framework can also be extended to other risk measures, including conditional value-at-risk (CVaR), without modification of the underlying Bayesian inference procedure.

The subsequent sections describe each component of the framework in detail. Section~\ref{sec:failure_probability} reformulates structural reliability analysis as a bounded expectation suitable for amplitude estimation. Section~\ref{sec:oracle} presents the oracle construction used throughout this work. Sections~\ref{sec:iqae} and \ref{sec:bayesian_iqae} introduce the iterative quantum amplitude estimation algorithm and the Bayesian updating procedure, respectively. 

\section{Structural Failure Probability as a Bounded Expectation}
\label{sec:failure_probability}

The Bayesian IQAE framework developed in this work relies on reformulating structural reliability analysis as the estimation of a bounded expectation. This representation establishes a direct connection between failure-probability estimation and quantum amplitude estimation.
Let $I(\boldsymbol{\xi})$ denote the binary failure indicator associated with the limit-state function introduced in \eref{eq:framework_limit_state}
\begin{equation}
I(\boldsymbol{\xi})
=
\begin{cases}
1, & g(\boldsymbol{\xi}) < 0,\\
0, & g(\boldsymbol{\xi}) \ge 0.
\end{cases}
\label{eq:failure_indicator}
\end{equation}
The corresponding failure probability defined in \eref{eq:framework_pf} can be written as
\begin{equation}
p_f
=
\mathbb{E}\!\left[I(\boldsymbol{\xi})\right].
\label{eq:pf_expectation}
\end{equation}
For a discrete stochastic ensemble consisting of $N_s$ realizations with
associated probabilities $p_i$, Eq.~\eqref{eq:pf_expectation} becomes
\begin{equation}
p_f
=
\sum_{i=1}^{N_s}
p_i I_i,
\label{eq:pf_discrete}
\end{equation}
where
$I_i = I(\boldsymbol{\xi}_i)$.
For uniformly weighted realizations, $p_i=1/N_s$, Eq.~\eqref{eq:pf_discrete}
reduces to
\begin{equation}
p_f
=
\frac{1}{N_s}
\sum_{i=1}^{N_s}
I_i.
\label{eq:pf_uniform}
\end{equation}

Equation~\eqref{eq:pf_expectation} reveals that structural reliability
analysis can be viewed as the estimation of a bounded expectation.
Because the failure indicator satisfies
\begin{equation}
0 \le I(\boldsymbol{\xi}) \le 1,
\label{eq:indicator_bounds}
\end{equation}
the target quantity is naturally confined to the interval $[0,1]$. This property is particularly important in the context of quantum amplitude estimation, where the target quantity must ultimately be represented as a probability amplitude. Because the failure indicator is already bounded between zero and one, the failure probability can be encoded directly as a quantum amplitude without requiring any additional scaling or normalization.

This property distinguishes failure probability from more general reliability and risk measures. Quantities such as expected structural response, tail expectations, or conditional value-at-risk (CVaR) typically involve continuous response variables that are not naturally confined to the interval $[0,1]$. In such cases, an additional normalization step is required before the quantity can be represented within an amplitude-estimation framework. Failure probability avoids this complication because the binary indicator function already satisfies the boundedness requirements imposed by quantum amplitude estimation. Consequently, structural reliability analysis provides a particularly natural setting for the Bayesian IQAE methodology developed in this work.

\section{Quantum Oracle Construction}
\label{sec:oracle}

The formulation developed in the previous section expresses the structural failure probability as the expectation of a binary-valued random variable. To leverage quantum amplitude estimation, this binary information must be encoded into a quantum oracle capable of distinguishing failed and safe realizations within a quantum superposition of stochastic states.

Consider a discrete stochastic ensemble consisting of $N_s$ realizations
$\{\boldsymbol{\xi}_i\}_{i=0}^{N_s-1}$. For each realization, the limit-state function is evaluated and the corresponding failure indicator is defined as
\begin{equation}
I_i = I(\boldsymbol{\xi}_i),
\qquad
I_i \in \{0,1\},
\label{eq:oracle_indicator}
\end{equation}
where $I_i=1$ denotes a failed realization and $I_i=0$ denotes a safe realization. The role of the quantum oracle is to identify and mark the failed realizations within the stochastic ensemble, thereby transforming the reliability problem into an amplitude-estimation problem.

Let $n=\lceil \log_2 N_s \rceil$ denote the number of qubits required to represent the stochastic ensemble. The computational basis state $\ket{i}$ is used to represent realization $\boldsymbol{\xi}_i$. A quantum state corresponding to the stochastic ensemble can be written as
\begin{equation}
\ket{\psi}
=
\sum_{i=0}^{N_s-1}
\sqrt{p_i}\ket{i},
\label{eq:stochastic_state}
\end{equation}
where $p_i$ denotes the probability associated with realization $i$, with
\begin{equation}
\sum_{i=0}^{N_s-1} p_i = 1.
\label{eq:probability_normalization}
\end{equation}
For uniformly weighted realizations, Eq.~\eqref{eq:stochastic_state} reduces to
\begin{equation}
\ket{\psi}
=
\frac{1}{\sqrt{N_s}}
\sum_{i=0}^{N_s-1}
\ket{i}.
\label{eq:uniform_state}
\end{equation}

An ancilla qubit is introduced to store the failure information. The oracle $U_f$ acts as
\begin{equation}
U_f:
\ket{i}\ket{0}
\longrightarrow
\ket{i}\ket{I_i},
\label{eq:oracle_action}
\end{equation}
thereby encoding the binary failure indicator into the ancilla register. Applying $U_f$ to the stochastic state gives
\begin{equation}
U_f
\left(
\sum_{i=0}^{N_s-1}
\sqrt{p_i}
\ket{i}\ket{0}
\right)
=
\sum_{i=0}^{N_s-1}
\sqrt{p_i}
\ket{i}\ket{I_i}.
\label{eq:oracle_state}
\end{equation}
Because $I_i$ is binary, the state in Eq.~\eqref{eq:oracle_state} can be separated into safe and failed components as
\begin{equation}
\ket{\Psi}
=
\sum_{I_i=0}
\sqrt{p_i}\ket{i}\ket{0}
+
\sum_{I_i=1}
\sqrt{p_i}\ket{i}\ket{1}.
\label{eq:safe_failed_decomposition}
\end{equation}
The probability of measuring the ancilla qubit in the state $\ket{1}$ is therefore obtained by summing the squared amplitudes of all basis states whose ancilla is $\ket{1}$
\begin{equation}
a
=
\sum_{I_i=1}
\left|\sqrt{p_i}\right|^2
=
\sum_{I_i=1}
p_i.
\label{eq:ancilla_one_probability}
\end{equation}
Equivalently, since $I_i=1$ for failed realizations and $I_i=0$ for safe realizations, this probability can be written as
\begin{equation}
a
=
\sum_{i=0}^{N_s-1}
p_i I_i
=
p_f,
\label{eq:oracle_amplitude}
\end{equation}
where the second equality follows directly from the discrete definition of the failure probability in Eq.~\eqref{eq:pf_discrete}.

To make the connection with amplitude estimation explicit, the post-oracle state in Eq.~\eqref{eq:oracle_state} can be rewritten as

\begin{equation}
\ket{\Psi}
=
\sqrt{1-a}\,\ket{\Psi_0}
+
\sqrt{a}\,\ket{\Psi_1},
\label{eq:ae_state_decomposition}
\end{equation}

where

\begin{equation}
\ket{\Psi_0}
=
\frac{1}{\sqrt{1-a}}
\sum_{I_i=0}
\sqrt{p_i}\,
\ket{i}\ket{0},
\label{eq:safe_state}
\end{equation}

and

\begin{equation}
\ket{\Psi_1}
=
\frac{1}{\sqrt{a}}
\sum_{I_i=1}
\sqrt{p_i}\,
\ket{i}\ket{1}.
\label{eq:failure_state}
\end{equation}

The state $\ket{\Psi_0}$ is the normalized superposition of all safe realizations, for which the ancilla qubit is in the state $\ket{0}$, while $\ket{\Psi_1}$ is the normalized superposition of all failed realizations, for which the ancilla qubit is in the state $\ket{1}$. The coefficient $\sqrt{a}$ is therefore the quantum amplitude associated with the failure subspace. Its squared magnitude
$|\sqrt{a}|^2 = a,$
is the probability of measuring the ancilla qubit in the state $\ket{1}$. Since Eq.~\eqref{eq:oracle_amplitude} establishes that $a=p_f$, estimation of the failure probability is equivalent to estimation of the amplitude associated with the failure subspace. Iterative quantum amplitude estimation exploits this representation to estimate $a$ using repeated applications of the oracle and Grover operator.
The reliability problem is therefore transformed into the estimation of the amplitude associated with the failure subspace. Once the oracle has been constructed, iterative quantum amplitude estimation can be applied directly to estimate the unknown failure probability.

In the present work, the oracle is implemented using a precomputed database of stochastic realizations and their associated failure indicators. This lookup-table representation allows the focus to remain on the statistical inference aspects of quantum reliability analysis while avoiding the substantially more challenging problem of constructing reversible finite-element solvers within a quantum circuit. The resulting oracle provides exact access to the failure information associated with each realization and serves as the foundation for the Bayesian IQAE framework developed in the following sections.
A detailed gate-level realization of the lookup-table oracle, including the state-preparation circuit and controlled-rotation implementation, is presented in Ref.~\cite{tabarraei2026stabilized}.

\subsection{Motivation for Quantum Amplitude Estimation}
\label{sec:computational_complexity}

The primary motivation for employing quantum amplitude estimation in reliability analysis is its improved sampling efficiency relative to classical Monte Carlo simulation. Since structural failure probability is represented by the amplitude $a=p_f$, estimation accuracy can be analyzed in terms of the number of oracle evaluations required to achieve a prescribed error tolerance.
For a classical Monte Carlo simulation, the failure probability is estimated from repeated observations of the binary failure indicator. Let $\widehat{p}_f^{\mathrm{MC}}$ denote the Monte Carlo estimator obtained from $N$ independent samples. The variance of the estimator is
\begin{equation}
\mathrm{Var}
\left(
\widehat{p}_f^{\mathrm{MC}}
\right)
=
\frac{p_f(1-p_f)}{N}.
\label{eq:mc_variance}
\end{equation}
Consequently, the root-mean-square error scales as
\begin{equation}
\mathrm{RMSE}_{\mathrm{MC}}
=
\mathcal{O}
\left(
N^{-1/2}
\right).
\label{eq:mc_rmse}
\end{equation}
To achieve an estimation accuracy $\epsilon$, the required number of samples therefore scales as
\begin{equation}
N_{\mathrm{MC}}
=
\mathcal{O}
\left(
\epsilon^{-2}
\right).
\label{eq:mc_complexity}
\end{equation}
This behavior becomes particularly problematic for rare-event reliability analysis because a large number of samples may be required before any failures are observed.

In contrast, quantum amplitude estimation exploits the amplitude-amplification process described in Section~\ref{sec:iqae}. By performing measurements at multiple Grover depths and combining the resulting information through statistical inference, the unknown amplitude can be estimated with asymptotically improved efficiency. For amplitude-estimation-based methods, the estimation error scales as
\begin{equation}
\mathrm{RMSE}_{\mathrm{QAE}}
=
\mathcal{O}
\left(
N^{-1}
\right),
\label{eq:qae_rmse}
\end{equation}
where $N$ denotes the total number of oracle evaluations. Equivalently, the number of oracle evaluations required to achieve accuracy $\epsilon$ satisfies
\begin{equation}
N_{\mathrm{QAE}}
=
\mathcal{O}
\left(
\epsilon^{-1}
\right).
\label{eq:qae_complexity}
\end{equation}
The quadratic improvement in oracle complexity represented by Eqs.~\eqref{eq:mc_complexity} and \eqref{eq:qae_complexity} is the principal theoretical advantage of amplitude-estimation-based reliability analysis. For increasingly stringent accuracy requirements, the gap between classical Monte Carlo and quantum amplitude estimation grows progressively larger.

 The Bayesian formulation introduced in the present work does not alter the asymptotic oracle complexity of iterative quantum amplitude estimation. The same quantum measurements are performed, and the same amplified probabilities govern the measurement process. The role of Bayesian inference is instead to provide a probabilistic characterization of the uncertainty associated with the estimated failure probability. The additional computational cost arises only from the classical posterior updates performed on the one-dimensional parameter grid introduced in Section~\ref{sec:posterior_representation}. Because the posterior is represented by a discretized distribution over a single scalar parameter, the cost of Bayesian updating is negligible compared with the cost of oracle evaluations and finite-element model evaluations.
Consequently, Bayesian IQAE retains the advantage of amplitude estimation while simultaneously providing posterior distributions, credible intervals, uncertainty-aware stopping criteria, and adaptive measurement strategies.

\section{Iterative Quantum Amplitude Estimation}
\label{sec:iqae}
\subsection{State Preparation and Amplitude Amplification}
The oracle construction developed in Section~\ref{sec:oracle} reduces the reliability problem to estimation of the parameter $a$, which is equal to the structural failure probability. Let $A$ denote the state-preparation operator that maps the initial quantum register to the state defined by Eq.~\eqref{eq:ae_state_decomposition}. Specifically
\begin{equation}
A\ket{0}^{\otimes n}\ket{0}
=
\sqrt{1-a}\,\ket{\Psi_0}
+
\sqrt{a}\,\ket{\Psi_1}.
\label{eq:state_preparation_operator}
\end{equation}
Without amplitude amplification, the state in Eq.~\eqref{eq:state_preparation_operator} can only be interrogated through direct measurement of the ancilla qubit. Because the probability of observing the outcome $\ket{1}$ is equal to $a$, repeated measurements yield a sequence of binary observations that is statistically equivalent to classical Monte Carlo sampling of the failure indicator. The resulting estimator therefore exhibits the standard Monte Carlo convergence rate of $\mathcal{O}(1/\sqrt{N})$, requiring $\mathcal{O}(1/\epsilon^2)$ samples to achieve an accuracy $\epsilon$. This becomes particularly inefficient for rare-event reliability analysis, where failures occur infrequently.

Iterative quantum amplitude estimation addresses this limitation through amplitude amplification. Since the states $\ket{\Psi_0}$ and $\ket{\Psi_1}$ correspond to disjoint safe and failure subspaces, they are orthonormal and form a basis for a two-dimensional subspace of the Hilbert space. Consequently, the state in Eq.~\eqref{eq:state_preparation_operator} may be interpreted geometrically as a unit vector in the plane spanned by $\ket{\Psi_0}$ and $\ket{\Psi_1}$.

Introducing the parameterization
\begin{equation}
a=\sin^2(\theta),
\qquad
\theta\in[0,\pi/2],
\label{eq:theta_parameterization}
\end{equation}
allows Eq.~\eqref{eq:state_preparation_operator} to be expressed as
\begin{equation}
A\ket{0}^{\otimes n}\ket{0}
=
\cos(\theta)\ket{\Psi_0}
+
\sin(\theta)\ket{\Psi_1}.
\label{eq:state_theta}
\end{equation}
In this representation, the unknown failure probability is encoded through the angle $\theta$. Amplitude amplification acts as a sequence of rotations within the two-dimensional subspace spanned by $\ket{\Psi_0}$ and $\ket{\Psi_1}$, increasing the component of the state associated with the failure subspace prior to measurement. The amplification process is governed by the Grover operator
\begin{equation}
\mathcal{G}=-AS_0A^\dagger S_{\chi},
\label{eq:grover_operator}
\end{equation}
where $A^\dagger$ denotes the Hermitian adjoint of the state-preparation operator $A$
\begin{equation}
S_0
=
I
-
2\ket{\mathbf{0}}\bra{\mathbf{0}},
\qquad
\ket{\mathbf{0}}
=
\ket{0}^{\otimes n}\ket{0},
\label{eq:s0_operator}
\end{equation}
is a reflection about the initial state, and
\begin{equation}
S_{\chi}
=
I
-
2\ket{\Psi_1}\bra{\Psi_1},
\label{eq:sf_operator}
\end{equation}
is a reflection about the failure subspace. Since the Grover operator is the product of two reflections, it acts as a rotation within the two-dimensional subspace spanned by $\ket{\Psi_0}$ and $\ket{\Psi_1}$. For brevity, the gate-level realization of the oracle, the reflection operators, and the Grover circuit is omitted here; the corresponding circuit implementation can be found in Ref.~\cite{tabarraei2026stabilized}.

After $k$ applications of the Grover operator, the quantum state becomes
\begin{equation}
\mathcal{G}^kA\ket{0}^{\otimes n}\ket{0}
=
\cos\bigl((2k+1)\theta\bigr)\ket{\Psi_0}
+
\sin\bigl((2k+1)\theta\bigr)\ket{\Psi_1}.
\label{eq:grover_rotation}
\end{equation}
The probability of observing the ancilla qubit in the state $\ket{1}$ is therefore
\begin{equation}
p_k=\sin^2\!\bigl((2k+1)\theta\bigr).
\label{eq:amplified_probability}
\end{equation}
For the special case $k=0$, Eq.~\eqref{eq:amplified_probability} reduces to
$p_0=\sin^2(\theta)=a$,
which corresponds exactly to direct Monte Carlo sampling of the oracle state. The use of $k>0$ amplifies the probability of observing the failure subspace and therefore provides more informative measurements for estimating rare-event probabilities.

\subsection{Statistical Measurement Model}
For a given Grover depth $k_i$, the amplified probability of observing the ancilla qubit in the state $\ket{1}$ is determined by Eq.~\eqref{eq:amplified_probability}. To estimate the unknown parameter $\theta$, the quantum circuit is executed repeatedly at the selected Grover depth. Suppose that $m_i$ shots are performed and $h_i$ successful outcomes are observed, where a success corresponds to measurement of the ancilla qubit in the state $\ket{1}$. The data collected at the $i$th Grover depth are therefore represented by
$D_i=(k_i,m_i,h_i).$
Because each circuit execution produces a binary outcome, the number of successes follows a binomial distribution
\begin{equation}
h_i
\sim
\mathrm{Binomial}
\!\left(
m_i,
p_{k_i}(\theta)
\right),
\label{eq:binomial_model_single}
\end{equation}
where $p_{k_i}(\theta)$ is the amplified success probability given by Eq.~\eqref{eq:amplified_probability}. The probability of observing the measurement batch $D_i$ for a given value of $\theta$ is therefore

\begin{equation}
p(D_i|\theta)
=
\binom{m_i}{h_i}
p_{k_i}(\theta)^{h_i}
\left(
1-p_{k_i}(\theta)
\right)^{m_i-h_i}.
\label{eq:binom_likelihood_single}
\end{equation}
Iterative quantum amplitude estimation proceeds by collecting measurement data from a sequence of Grover depths. After $t$ measurement rounds, the complete dataset is
\begin{equation}
 D_t
=
\{D_i\}_{i=1}^{t}
=
\{(k_i,m_i,h_i)\}_{i=1}^{t}.   
\end{equation}
Assuming conditional independence between measurement batches, the likelihood of the entire dataset is obtained by multiplying the individual likelihood contributions
\begin{equation}
p(D_t|\theta)
=
\prod_{i=1}^{t}
\binom{m_i}{h_i}
p_{k_i}(\theta)^{h_i}
\left(
1-p_{k_i}(\theta)
\right)^{m_i-h_i}.
\label{eq:likelihood_full}
\end{equation}

\subsection{Construction of the Admissible Amplitude Interval}
\label{sec:admissible_interval}

The statistical model developed in the previous subsection provides the
likelihood of the observed quantum measurements for any candidate
amplitude angle $\theta$. However, because the amplified success
probability is related to $\theta$ through the periodic transformation
given in Eq.~\eqref{eq:amplified_probability}, a single set of
measurement outcomes may be consistent with multiple values of
$\theta$, particularly at larger Grover depths. Consequently, before
constructing a Bayesian posterior, it is necessary to identify the
subset of amplitude angles that remain statistically compatible with the
observed measurements. This admissible region forms the basis of the
interval-based IQAE algorithm of Grinko et al.~\cite{Grinko2021} and is
also adopted in the present Bayesian formulation.

Following the interval-based IQAE framework of Grinko et al.~\cite{Grinko2021},
the present implementation constructs this range using the exact
Clopper--Pearson confidence interval for a binomial proportion~\cite{clopper1934use}.
For a confidence level $1-\alpha$, the lower and upper limits for the amplified
success probability are denoted by $p_L^{(t)}$ and $p_U^{(t)}$, respectively.
They are obtained from the beta distribution quantiles as
\begin{equation}
p_L^{(t)} =
\begin{cases}
0, & h_t = 0,\\
B^{-1}\!\left(\alpha/2;\, h_t,\, m_t-h_t+1\right), & h_t > 0,
\end{cases}
\label{eq:cp_lower}
\end{equation}
and
\begin{equation}
p_U^{(t)} =
\begin{cases}
1, & h_t = m_t,\\
B^{-1}\!\left(1-\alpha/2;\, h_t+1,\, m_t-h_t\right), & h_t < m_t,
\end{cases}
\label{eq:cp_upper}
\end{equation}
where $B^{-1}(q;\,a,b)$ denotes the $q$th quantile of the beta distribution with
shape parameters $a$ and $b$. The resulting interval
$[p_L^{(t)},p_U^{(t)}]$ gives the set of amplified success probabilities that are
statistically compatible with the observed number of successes at iteration $t$.

The admissible interval for the amplified probability is then mapped back
to the amplitude-angle domain using the amplification relation in
Eq.~\eqref{eq:amplified_probability}. Because this mapping is periodic in
$\theta$, a single confidence interval for the amplified probability may
correspond to multiple disjoint intervals in the amplitude angle. The
admissible set associated with the $t$th measurement batch is therefore
\begin{equation}
\Theta_t=
\left\{
\theta\in[0,\pi/2]:
p_L^{(t)}
\le
\sin^2\!\big((2k_t+1)\theta\big)
\le
p_U^{(t)}
\right\}.
\label{eq:theta_admissible_batch}
\end{equation}

Each connected component of $\Theta_t$ represents a candidate branch of
the amplitude angle that is statistically consistent with the current
measurement batch. As additional measurements are collected at different
Grover depths, the admissible branches are updated according to the new
measurement information. Branches that become incompatible with the
observed data are discarded, while the remaining branches continue to be
refined. Consequently, the number of admissible branches typically
decreases as the IQAE procedure progresses, leading to a progressively
more localized estimate of the amplitude.

In the proposed Bayesian formulation, the admissible branches define the
feasible region over which inference is performed. Rather than treating
all surviving branches equally, the Bayesian posterior introduced in the next section assigns relative
probability mass to each branch according to the accumulated likelihood
from all measurement batches. As new observations are incorporated,
posterior probability shifts toward the branches that are most strongly
supported by the data while probability mass associated with less
plausible branches diminishes. The resulting posterior distribution
therefore provides both a point estimate of the failure probability and a
rigorous quantification of the remaining uncertainty.

\subsection{Maximum-Likelihood IQAE}
For completeness and to facilitate comparison with the proposed Bayesian formulation, we briefly review the maximum-likelihood version of IQAE, which serves as the classical inference procedure for estimating the unknown amplitude from quantum measurement data. The maximum-likelihood formulation estimates the unknown parameter $\theta$ by identifying the value that maximizes the likelihood function in Eq.~\eqref{eq:likelihood_full}. Specifically
\begin{equation}
\widehat{\theta}_{\mathrm{MLE}}
=
\arg\max_{\theta\in\Theta^{(t)}}
p(D_t|\theta).
\label{eq:mle_theta}
\end{equation}
The corresponding estimate of the failure probability is then obtained through the relationship between $\theta$ and the amplitude
\begin{equation}
\widehat{a}_{\mathrm{MLE}}
=
\sin^2
\!\left(
\widehat{\theta}_{\mathrm{MLE}}
\right).
\label{eq:mle_amplitude}
\end{equation}
The maximum-likelihood formulation provides a point estimate of the unknown amplitude but does not directly quantify the uncertainty associated with the estimate. Furthermore, for rare-event problems and limited measurement budgets, the likelihood function may remain relatively broad, making uncertainty quantification particularly important. These limitations motivate the Bayesian formulation developed in the next section, where the likelihood in Eq.~\eqref{eq:likelihood_full} is combined with a prior distribution to construct a posterior distribution for the failure probability.

The binomial coefficient in Eq.~\eqref{eq:likelihood_full} is independent of $\theta$ and therefore does not affect either the maximum-likelihood estimate or the normalized Bayesian posterior. Consequently, it may be retained in the formal probabilistic model while being omitted from numerical likelihood evaluations for computational efficiency.

\section{Bayesian Iterative Quantum Amplitude Estimation}
\label{sec:bayesian_iqae}

The IQAE measurement model developed in Section~\ref{sec:iqae} provides a direct statistical link between quantum circuit measurements and the unknown failure probability. In maximum-likelihood IQAE, this information is compressed into a single point estimate. While such an estimate is useful, reliability analysis requires more than a point prediction. In rare-event problems, the central question is not only what the estimated failure probability is, but also how much uncertainty remains in that estimate after a finite number of quantum measurements. This distinction is especially important when the number of observed failures is small, the likelihood is broad, or multiple values of the amplitude remain statistically plausible.

The Bayesian formulation developed in this section addresses this issue by treating the unknown angle $\theta$ as a random variable and updating its probability distribution as measurement data are collected. Instead of returning only a single estimate, Bayesian IQAE produces a posterior distribution over $\theta$ and, through the transformation $a=\sin^2(\theta)$, a posterior distribution over the failure probability. This posterior distribution provides a complete uncertainty-aware description of the reliability estimate, including posterior means, credible intervals, and uncertainty measures that can be monitored throughout the algorithm.

\subsection{Bayesian Posterior Construction}
\label{sec:posterior_construction}

The maximum-likelihood formulation discussed in the previous section identifies the single value of $\theta$ that maximizes the likelihood function. While this approach provides a point estimate, it does not directly quantify the uncertainty associated with the estimate. In reliability analysis, particularly for rare-event problems, uncertainty quantification is often as important as the estimate itself. Bayesian IQAE addresses this limitation by treating $\theta$ as an uncertain parameter and updating its probability distribution as measurement data are collected.

Let $\pi_0(\theta)$ denote the prior distribution assigned to $\theta$ before any measurements are observed. In the absence of prior information, a uniform prior over the admissible interval is adopted,

\begin{equation}
\pi_0(\theta)
=
\frac{2}{\pi},
\qquad
\theta\in[0,\pi/2].
\label{eq:theta_prior_uniform}
\end{equation}

Given the measurement dataset $D_t$ and the likelihood function defined in Eq.~\eqref{eq:likelihood_full}, Bayes' theorem yields the posterior distribution
\begin{equation}
\pi_t(\theta)
=
p(\theta|D_t)
=
\frac{p(D_t|\theta)\,\pi_0(\theta)}
{\int_0^{\pi/2} p(D_t|\vartheta)\,\pi_0(\vartheta)\,d\vartheta},
\label{eq:bayes_posterior}
\end{equation}
where $\vartheta$ is a dummy integration variable. The denominator serves as a normalization constant that ensures the posterior integrates to unity over the admissible parameter space.

Equation~\eqref{eq:bayes_posterior} forms the foundation of Bayesian IQAE. The likelihood function quantifies how well a candidate value of $\theta$ explains the observed quantum measurements, while the prior represents the information available before any measurements are collected. The posterior combines these two sources of information into a single probability distribution that characterizes the remaining uncertainty in $\theta$ after observing the data.

This posterior is the central object in Bayesian IQAE. Each measurement batch reshapes the distribution by increasing the probability density near values of $\theta$ that are consistent with the observed amplified success counts and decreasing the probability density near values that are inconsistent with the data. As the algorithm progresses, the posterior contracts around the values of $\theta$ supported by the quantum measurements.

\subsection{Sequential Bayesian Updating}
\label{sec:sequential_update}

A key advantage of the Bayesian formulation is that inference can be performed sequentially as new IQAE measurements become available. Rather than recomputing the posterior distribution from the complete dataset after each measurement round, the posterior obtained from previous measurements can be used directly as the prior for the next update. In this manner, information is accumulated recursively throughout the IQAE process.

Let $\pi_{t-1}(\theta)$ denote the posterior distribution after processing the first $t-1$ measurement batches. When a new measurement batch $D_t$ is acquired, Bayes' theorem updates the posterior according to

\begin{equation}
\pi_t(\theta)
=
\frac{p(D_t|\theta)\,\pi_{t-1}(\theta)}
{\int_0^{\pi/2}
p(D_t|\phi)\,
\pi_{t-1}(\phi)\,
d\phi},
\label{eq:recursive_posterior}
\end{equation}

where the likelihood function is given by Eq.~\eqref{eq:binom_likelihood_single}. Substituting the binomial likelihood into Eq.~\eqref{eq:recursive_posterior} yields

\begin{equation}
\pi_t(\theta)
=
\frac{
p_{k_t}(\theta)^{h_t}
\left[1-p_{k_t}(\theta)\right]^{m_t-h_t}
\pi_{t-1}(\theta)}
{\int_0^{\pi/2}
p_{k_t}(\phi)^{h_t}
\left[1-p_{k_t}(\phi)\right]^{m_t-h_t}
\pi_{t-1}(\phi)\,
d\phi},
\label{eq:recursive_binomial_update}
\end{equation}

where the binomial coefficient has been omitted because it is independent of $\theta$ and therefore cancels during posterior normalization.

Equation~\eqref{eq:recursive_binomial_update} shows explicitly how each new measurement batch modifies the current state of knowledge. Values of $\theta$ that predict amplified success probabilities consistent with the observed measurements receive increased posterior weight, whereas values that are inconsistent with the data are progressively suppressed. As additional measurements are collected, the posterior distribution contracts around the region of parameter space supported by the accumulated quantum evidence. This posterior contraction is the statistical mechanism through which Bayesian IQAE converts quantum measurements into uncertainty reduction. Informative measurements produce rapid contraction and narrow posterior distributions, while less informative measurements lead to broader distributions that reflect the remaining uncertainty in the inferred amplitude.

The sequential structure of Eq.~\eqref{eq:recursive_binomial_update} is particularly attractive for adaptive IQAE. Because the posterior distribution is updated immediately after each measurement batch, the current uncertainty can be used to guide future measurement decisions, including the selection of Grover depths, shot allocations, and stopping criteria. The posterior therefore serves both as an estimator of the unknown failure probability and as a quantitative measure of the information gained during the inference process.

Although posterior contraction generally reduces uncertainty, the relationship between the measurements and the unknown parameter is not always one-to-one. An important feature of IQAE is that the amplified success probability is governed by the periodic function $\sin^2((2k+1)\theta)$. Consequently, different values of $\theta$ may occasionally produce similar measurement statistics, particularly when large Grover depths are employed. In such situations, the posterior distribution may become multimodal, indicating that multiple parameter values remain consistent with the observed data. Rather than concealing this ambiguity through a single point estimate, the Bayesian formulation represents it explicitly through the posterior distribution.

This ability to quantify uncertainty is one of the principal advantages of Bayesian IQAE. Whereas maximum-likelihood approaches reduce the inference problem to a single estimate, Bayesian IQAE maintains the full distribution of plausible parameter values throughout the estimation process. The posterior therefore provides both an estimate of the failure probability and a quantitative characterization of the uncertainty associated with that estimate.

\subsection{Posterior Statistics and Uncertainty Quantification}
\label{sec:posterior_statistics}

The posterior distribution obtained through Bayesian updating contains substantially more information than a single point estimate. While maximum-likelihood IQAE returns only the most likely value of the unknown parameter, Bayesian IQAE provides a complete probabilistic description of the remaining uncertainty after a finite number of quantum measurements. This distinction is particularly important in reliability analysis, where the confidence associated with an estimated failure probability is often as important as the estimate itself.

The posterior distribution is constructed over the parameter $\theta$. However, the quantity of primary interest is the failure probability. Using the relationship established in Eq.~\eqref{eq:theta_parameterization}, the posterior distribution over $\theta$ induces a corresponding posterior distribution over the failure probability. Consequently, any posterior statistic of interest can be computed directly from the posterior distribution obtained through Eqs.~\eqref{eq:bayes_posterior} and \eqref{eq:recursive_binomial_update}.

A natural Bayesian estimate of the failure probability is the posterior mean
\begin{equation}
\widehat{a}_{\mathrm{B}}
=
\int_{0}^{\pi/2}
\sin^2(\theta)\,
\pi_t(\theta)\,
d\theta,
\label{eq:posterior_mean_amplitude}
\end{equation}
which represents the expected value of the failure probability under the posterior distribution. Unlike the maximum-likelihood estimate, which identifies a single most probable value, the posterior mean incorporates information from the entire posterior distribution and therefore reflects both the location and shape of the inferred probability density.
The uncertainty associated with the estimate may be quantified through the posterior variance
\begin{equation}
\mathrm{Var}_{\mathrm{B}}(a)
=
\int_{0}^{\pi/2}
\left[
\sin^2(\theta)
-
\widehat{a}_{\mathrm{B}}
\right]^2
\pi_t(\theta)\,
d\theta,
\label{eq:posterior_variance_amplitude}
\end{equation}
and the corresponding posterior standard deviation
\begin{equation}
\sigma_{\mathrm{B}}(a)
=
\sqrt{\mathrm{Var}_{\mathrm{B}}(a)}.
\label{eq:posterior_std_amplitude}
\end{equation}

For rare-event reliability problems, the magnitude of the uncertainty is often more informative when expressed relative to the estimated failure probability. A convenient normalized measure is the posterior coefficient of variation
\begin{equation}
\mathrm{CV}_{\mathrm{B}}(a)
=
\frac{\sigma_{\mathrm{B}}(a)}
{\widehat{a}_{\mathrm{B}}},
\label{eq:posterior_cv}
\end{equation}
which provides a dimensionless measure of the remaining uncertainty in the estimate. As additional measurement batches are incorporated through the sequential update process, the posterior distribution contracts and the coefficient of variation decreases. Consequently, $\mathrm{CV}_{\mathrm{B}}(a)$ provides a useful indicator of convergence and estimation confidence.

In addition to moment-based measures, Bayesian inference naturally provides credible intervals that quantify the range of failure probabilities supported by the posterior distribution. Let $a_q$ denote the $q$th posterior quantile of the failure probability $a=\sin^2(\theta)$. The equal-tail $95\%$ credible interval is defined as
\begin{equation}
CI_{95}
=
\left[
a_{0.025},
a_{0.975}
\right].
\label{eq:credible_interval}
\end{equation}
The interval $CI_{95}$ contains $95\%$ of the posterior probability mass associated with the unknown failure probability. The corresponding credible-interval width is
\begin{equation}
\Delta CI_{95}
=
a_{0.975}
-
a_{0.025}.
\label{eq:credible_interval_width}
\end{equation}

The width of the credible interval provides a direct measure of the uncertainty remaining in the estimated failure probability. Broad intervals indicate that the available quantum measurements are consistent with a wide range of failure probabilities, whereas narrow intervals indicate strong posterior concentration and increased confidence in the estimate. Unlike point estimators, which summarize the posterior distribution by a single value, credible intervals explicitly quantify the range of plausible failure probabilities supported by the data.

For this reason, credible intervals play a central role in the present work and are used throughout the numerical examples to assess posterior uncertainty, monitor convergence, and compare Bayesian IQAE with classical Monte Carlo simulation and maximum-likelihood IQAE.
The posterior mean, posterior variance, coefficient of variation, and credible intervals together provide a comprehensive uncertainty-aware characterization of the failure probability. These quantities form the basis for the adaptive measurement strategies and stopping criteria developed in the following sections.

\section{Computational Framework}
\label{sec:computational_framework}
The previous sections established the Bayesian formulation of iterative quantum amplitude estimation and the corresponding posterior update equations. This section describes the numerical implementation used throughout the present work, including posterior representation, uncertainty quantification, numerical stabilization, and the overall Bayesian IQAE workflow.

\subsection{Posterior Representation}
\label{sec:posterior_representation}

The posterior distribution is represented on a discrete grid spanning the admissible parameter interval $\theta\in[0,\pi/2]$. Let $\theta_j$, $j=1,\ldots,N_\theta$, denote the grid points used to discretize the parameter space, where $N_\theta$ is the total number of grid points. Because IQAE involves inference over a single scalar parameter, a grid-based representation provides an accurate and computationally efficient alternative to sampling-based approaches such as Markov chain Monte Carlo. The posterior distribution is evaluated and updated at each grid point according to Eqs.~\eqref{eq:bayes_posterior} and \eqref{eq:recursive_binomial_update}.

The posterior distribution is initialized using the prior described in Eq.~\eqref{eq:theta_prior_uniform}. As measurement batches are acquired, posterior probabilities are updated sequentially over the discretized parameter space. At each iteration, the posterior is normalized to ensure that it integrates to unity. The resulting grid-based representation provides a complete approximation of the posterior distribution and enables direct computation of posterior means, variances, credible intervals, and other uncertainty measures without the need for additional sampling procedures.

\subsection{Numerical Stabilization}
\label{sec:numerical_stabilization}

The posterior distribution is evaluated on the discretized parameter grid introduced in Section~\ref{sec:posterior_representation}. At each grid point $\theta_j$, the posterior is updated using the measurement data accumulated during the IQAE procedure. In principle, the posterior could be computed directly from the recursive update in Eq.~\eqref{eq:recursive_binomial_update}. However, this approach becomes numerically unstable when many measurement batches are incorporated.

The difficulty arises because the posterior is formed from a product of likelihood contributions. Each likelihood term is a probability between zero and one. As additional measurement batches are collected, repeated multiplication of these small numbers may produce values that are smaller than machine precision. This phenomenon, known as numerical underflow, can occur even when the relative shape of the posterior remains physically meaningful. The problem becomes particularly severe when the posterior distribution is sharply concentrated, which is precisely the regime of interest because it corresponds to highly informative quantum measurements.

To avoid numerical underflow, posterior updates are performed in logarithmic form. Instead of multiplying likelihood contributions, their logarithms are accumulated as sums. For the $i$th measurement batch, the log-likelihood contribution is
\begin{equation}
\ell_i(\theta)
=
h_i\log p_{k_i}(\theta)
+
(m_i-h_i)
\log\!\left(
1-p_{k_i}(\theta)
\right),
\label{eq:log_likelihood_batch}
\end{equation}
where terms independent of $\theta$ have been omitted because they do not affect the normalized posterior distribution. 
The cumulative log-posterior after $t$ measurement batches is then
\begin{equation}
\log \widetilde{\pi}_t(\theta)
=
\log \pi_0(\theta)
+
\sum_{i=1}^{t}
\ell_i(\theta),
\label{eq:cumulative_log_posterior}
\end{equation}
where $\widetilde{\pi}_t(\theta)$ denotes the unnormalized posterior density. Equation~\eqref{eq:cumulative_log_posterior} is evaluated independently at every grid point $\theta_j$. Thus, the Bayesian update reduces to the accumulation of log-likelihood contributions over the parameter grid as new IQAE measurements become available.

After the log-posterior has been computed, the posterior density is recovered through exponentiation. To improve numerical stability further, the maximum log-posterior value is first subtracted from all grid points
\begin{equation}
\widetilde{\pi}_t(\theta_j)
=
\exp
\!\left[
\log \widetilde{\pi}_t(\theta_j)
-
\max_k
\log \widetilde{\pi}_t(\theta_k)
\right].
\label{eq:stable_exponentiation}
\end{equation}
This operation preserves the relative shape of the posterior while preventing overflow during exponentiation. The posterior is then normalized according to
\begin{equation}
\pi_t(\theta_j)
=
\frac{\widetilde{\pi}_t(\theta_j)}
{\sum_{k=1}^{N_\theta}
\widetilde{\pi}_t(\theta_k)\Delta\theta},
\label{eq:grid_normalization}
\end{equation}
where $\Delta\theta$ denotes the grid spacing.
The normalized posterior obtained from Eq.~\eqref{eq:grid_normalization} is subsequently used to compute all posterior statistics, including the posterior mean, variance, coefficient of variation, and credible intervals. Consequently, the log-posterior formulation is not a modification of the Bayesian inference procedure itself, but rather a numerically stable implementation of the posterior updates required by Bayesian IQAE.

\subsection{Adaptive Grover-Depth and Shot Selection}
\label{sec:adaptive_k_m}

The efficiency of IQAE depends critically on the selection of the Grover depth and shot allocation at each iteration. The Grover depth controls the degree of amplitude amplification and therefore determines how informative the resulting measurements are. If the Grover depth is too small, the amplified success probability remains close to the original amplitude and the measurements provide limited information about the unknown parameter. Conversely, excessively large Grover depths may lead to ambiguity because the amplified success probability given by Eq.~\eqref{eq:amplified_probability} is periodic in $\theta$. As a result, distinct values of $\theta$ can produce nearly identical measurement statistics, making it difficult to distinguish between competing parameter estimates. The selection of $k_t$ must therefore balance two competing objectives: increasing amplification to improve measurement sensitivity while preserving sufficient identifiability of the unknown parameter.

Shot allocation introduces a similar tradeoff. Increasing the number of shots reduces statistical uncertainty in the observed success probability but consumes a larger fraction of the available computational budget. Since measurements performed at larger Grover depths are more expensive, the choice of shot count cannot be separated from the choice of amplification level. An effective IQAE strategy must therefore adapt both the Grover depth and shot allocation in a coordinated manner to maximize information gain while respecting the prescribed oracle-evaluation budget.

In the present implementation, both the Grover depth $k_t$ and the number of shots $m_t$ are selected adaptively using information contained in the current posterior distribution. Let
\begin{equation}
\widehat{\theta}_{t-1}
=
\int_{0}^{\pi/2}
\theta \,\pi_{t-1}(\theta)\, d\theta
\label{eq:posterior_mean_theta}
\end{equation}
denote the posterior mean after processing the first $t-1$ measurement batches. A nominal Grover depth is then proposed by placing the amplified angle near $\pi/4$, i.e.,
\begin{equation}
(2k_{\mathrm{prop}}+1)\widehat{\theta}_{t-1}\approx \frac{\pi}{4},
\end{equation}
which gives
\begin{equation}
k_{\mathrm{prop}}
=
\max\!\left\{
0,\;
\mathrm{round}\!\left(
\frac{1}{2}
\left(
\frac{\pi}{4\widehat{\theta}_{t-1}}-1
\right)
\right)
\right\}.
\label{eq:k_proposal}
\end{equation}

The motivation for Eq.~\eqref{eq:k_proposal} follows from the amplified success probability given by Eq.~\eqref{eq:amplified_probability}
whose sensitivity to $\theta$ is governed by
\begin{equation}
\frac{d p_k}{d\theta}
=
(2k+1)\sin\!\bigl(2(2k+1)\theta\bigr).
\label{eq:pk_theta_derivative}
\end{equation}
The magnitude of this derivative is largest when
\[
(2k+1)\theta \approx \frac{\pi}{4}\quad (\mathrm{mod}\ \pi/2),
\]
so selecting $k_{\mathrm{prop}}$ according to Eq.~\eqref{eq:k_proposal} places the expected amplified probability in a region where small changes in $\theta$ produce large changes in the measurement statistics, thereby increasing the information content of the next measurement batch.

The proposed Grover depth is not accepted unconditionally. Instead, it is subjected to a sequence of admissibility checks based on the current posterior distribution. Candidate values that introduce excessive aliasing, produce poor discrimination across the posterior support, or exceed a prescribed amplification limit are rejected. Additional safeguards restrict abrupt changes in Grover depth and ensure compatibility with the remaining computational budget. When posterior contraction stagnates, larger Grover depths may be introduced to increase amplification and accelerate information gain.

Shot allocation is performed jointly with Grover-depth selection. Because each Grover-amplified circuit requires approximately $2k_t+1$ oracle applications, the oracle cost associated with the $t$th measurement batch is
\begin{equation}
C_t=
m_t(2k_t+1),
\label{eq:batch_cost}
\end{equation}
where $m_t$ denotes the number of shots executed at depth $k_t$. Consequently, larger Grover depths provide stronger amplification but consume a larger fraction of the available computational budget.

The number of shots is selected adaptively to balance amplification strength and statistical precision. Additional shots are allocated when the posterior remains broad, when multiple posterior modes are present, or when convergence has stalled. Conversely, fewer shots are required once the posterior has become highly concentrated. The overall measurement process is constrained by the total oracle-evaluation budget
\begin{equation}
\sum_{t=1}^{N_{\mathrm{iter}}}
m_t(2k_t+1)
\leq
B,
\label{eq:budget_constraint}
\end{equation}
where $B$ denotes the prescribed oracle-evaluation budget.

The resulting adaptive strategy concentrates computational resources where they provide the greatest reduction in posterior uncertainty. Rather than prescribing a fixed sequence of Grover depths and shot counts, the algorithm continuously adjusts its measurement strategy in response to the evolving posterior distribution. This adaptive behavior is particularly advantageous for rare-event reliability estimation, where the amount of amplification and the number of measurements required to achieve a desired level of confidence may vary substantially from one problem to another.





\subsection{Stopping Criteria and Reliability-Oriented Error Measures}
\label{sec:stopping_criteria}

A distinguishing feature of Bayesian IQAE is that convergence can be assessed directly through the posterior distribution. Unlike conventional stopping rules based solely on the number of circuit evaluations, Bayesian stopping criteria quantify the remaining uncertainty in the failure probability estimate itself. This distinction is particularly important for rare-event reliability analysis, where identical measurement budgets may produce substantially different levels of confidence depending on the information content of the observed quantum measurements.

One natural uncertainty measure is the relative width of the $95\%$ posterior credible interval
\begin{equation}
\eta
=
\frac{\Delta CI_{95}}
{p_f}
=
\frac{a_{0.975}-a_{0.025}}
{p_f},
\label{eq:relative_credible_width}
\end{equation}
where $a_{0.025}$ and $a_{0.975}$ denote the lower and upper posterior quantiles, and $p_f$ denotes the reference failure probability. The quantity $\eta$ measures the posterior uncertainty relative to the magnitude of the failure probability and therefore provides a scale-independent measure of posterior concentration. In the numerical examples presented in this work, the reference failure probability is available and is therefore used to facilitate comparisons across different rarity levels. In practical applications where $p_f$ is unknown, the posterior mean estimate $\widehat{a}_{\mathrm{B}}$ may be used in its place.
A stopping criterion based on credible intervals is obtained by requiring
\begin{equation}
\eta
\leq
\varepsilon_{\mathrm{tol}},
\label{eq:credible_width_stopping}
\end{equation}
where $\varepsilon_{\mathrm{tol}}$ is a user-prescribed tolerance that specifies the maximum acceptable relative uncertainty in the failure probability estimate.

An alternative stopping metric is the posterior coefficient of variation introduced in Eq.~\eqref{eq:posterior_cv}. The IQAE procedure may be terminated when
\begin{equation}
\mathrm{CV}_{\mathrm{B}}(a)
\leq
\eta_{\mathrm{tol}},
\label{eq:cv_stopping}
\end{equation}
where $\eta_{\mathrm{tol}}$ denotes a prescribed threshold for the relative posterior standard deviation. Since the coefficient of variation measures uncertainty relative to the posterior mean, it provides a convenient dimensionless indicator of estimation confidence.

The criteria in Eqs.~\eqref{eq:credible_width_stopping} and \eqref{eq:cv_stopping} quantify convergence in terms of the reliability estimate rather than the computational effort expended to obtain it. As additional IQAE measurements are incorporated, the posterior distribution contracts, the credible interval narrows, and the coefficient of variation decreases. Consequently, these stopping criteria adapt naturally to the difficulty of the underlying estimation problem. Rare-event cases that require more information automatically demand additional measurements, whereas simpler cases terminate once the prescribed level of confidence has been achieved.


The individual components of the Bayesian IQAE framework described in the preceding sections are combined into the computational procedure summarized in Algorithm~\ref{alg:bayesian_iqae}. The algorithm begins with initialization of the prior distribution and computational budget, followed by adaptive selection of the Grover depth and shot allocation based on the current posterior distribution. Quantum measurements are then incorporated through Bayesian updating, and posterior statistics are evaluated to quantify the estimated failure probability and its associated uncertainty. The procedure continues until either a prescribed uncertainty tolerance is achieved or the available oracle-evaluation budget is exhausted.

\begin{algorithm}[!tbh]
\caption{Bayesian iterative quantum amplitude estimation for failure probability estimation}
\label{alg:bayesian_iqae}
\begin{algorithmic}[1]
\State Construct the state-preparation operator $A$ and Grover operator $\mathcal{G}$
\State Discretize the amplitude-angle domain $\theta\in[0,\pi/2]$ using grid points $\{\theta_j\}_{j=1}^{N_\theta}$
\State Initialize the prior distribution $\pi_0(\theta_j)$ on the grid
\State Set total oracle budget $B$, maximum Grover depth $K_{\max}$, shot bounds $m_{\min}$ and $m_{\max}$, confidence level $1-\alpha$, and stopping tolerance $\eta_{\mathrm{tol}}$
\State Set cumulative oracle cost $C_{\mathrm{tot}}=0$ and measurement dataset $D_0=\emptyset$
\For{$t=1,2,\ldots,N_{\mathrm{iter}}$}
\State Compute the current posterior summary, including the posterior mean or MAP estimate $\widehat{\theta}_{t-1}$ and the credible interval for $a$
\State Propose a nominal Grover depth
\[
k_{\mathrm{prop}}
=
\max\!\left\{
0,\,
\mathrm{round}
\!\left(
\frac{\pi}{8\max(\widehat{\theta}_{t-1},\theta_{\min})}
-
\frac{1}{2}
\right)
\right\}
\]
\State Select $k_t$ by applying posterior-support, aliasing, growth, $K_{\max}$, and remaining-budget constraints to $k_{\mathrm{prop}}$
\State Select the shot count $m_t$ using the current posterior uncertainty, shot bounds, and remaining budget
\State Execute the amplified circuit $\mathcal{G}^{k_t}A\ket{0}^{\otimes n}\ket{0}$ for $m_t$ shots
\State Record $h_t$, the number of measurements with ancilla outcome $\ket{1}$
\State Update the cumulative oracle cost
\[
C_{\mathrm{tot}}
\leftarrow
C_{\mathrm{tot}}
+
m_t(2k_t+1)
\]
\State Add the new measurement batch to the dataset:
\[
D_t \leftarrow D_{t-1}\cup\{(k_t,m_t,h_t)\}
\]
\State Compute the Clopper--Pearson interval $[p_L^{(t)},p_U^{(t)}]$ for the amplified success probability
\State Update the admissible branches in $\theta$ using the relation $p_{k_t}(\theta)=\sin^2((2k_t+1)\theta)$
\State Compute $p_{k_i}(\theta_j)$ for all measurement batches in $D_t$ and all grid points $\theta_j$
\State Update the log-posterior over the grid using the accumulated binomial log-likelihood
\[
\log\widetilde{\pi}_t(\theta_j)
\leftarrow
\log\pi_0(\theta_j)
+
\sum_{i=1}^{t}
\left[
h_i\log p_{k_i}(\theta_j)
+
(m_i-h_i)\log\!\left(1-p_{k_i}(\theta_j)\right)
\right]
\]
\State Stabilize the log-posterior by subtracting its maximum value over the grid
\State Exponentiate and normalize the posterior over the grid to obtain $\pi_t(\theta_j)$
\State Transform the posterior from $\theta$ to $a=\sin^2(\theta)$
\State Compute $\widehat{a}_{\mathrm{B}}$, posterior variance, $\mathrm{CV}_{\mathrm{B}}(a)$, and the credible interval for $a$
\If{$\mathrm{CV}_{\mathrm{B}}(a)\leq\eta_{\mathrm{tol}}$}
\State Terminate
\EndIf
\If{$C_{\mathrm{tot}}\geq B$}
\State Terminate
\EndIf
\EndFor
\State Return $\widehat{a}_{\mathrm{B}}$, posterior credible interval, posterior uncertainty measures, admissible-branch diagnostics, and total oracle cost
\end{algorithmic}
\end{algorithm}

\section{Numerical Results}
\label{sec:numerical_results}

This section evaluates the proposed Bayesian IQAE framework for rare-event structural reliability analysis. Two benchmark problems are considered. The first example is a stochastic one-dimensional bar problem, which serves as a controlled benchmark for studying the statistical behavior of the Bayesian IQAE estimator. The second example considers an L-shaped domain with an explicit quantum implementation of the lookup-table oracle, demonstrating the application of the proposed framework to a more realistic structural reliability problem.
For all examples, the Bayesian IQAE results are compared with classical Monte Carlo simulations using equivalent oracle budgets.  The comparison is made in an idealized oracle-query model, where one Monte Carlo sample and one invocation of the failure-marking oracle are treated as the primitive unit of cost. This isolates estimator efficiency from oracle-construction cost. Unless otherwise stated, all reported statistics are obtained from 30 independent runs using different random seeds.

For each benchmark problem, a fixed ensemble of stochastic realizations is generated prior to the reliability analysis. A finite-element simulation is performed for every realization, and the resulting response quantities are stored in a database. The corresponding failure indicators are subsequently evaluated using the prescribed limit-state function and encoded into a lookup-table oracle. The same stochastic database is used by Bayesian IQAE, maximum-likelihood IQAE, and classical Monte Carlo simulation, ensuring that all methods operate on an identical set of realizations and estimate the same underlying failure probability.

The proposed Bayesian IQAE framework is independent of the specific random-field representation adopted to generate the stochastic ensemble. Alternative uncertainty models, covariance kernels, stochastic discretization techniques, or sampling procedures may be employed without modification of the Bayesian inference procedure or the quantum amplitude-estimation framework. The stochastic-field model used in the present work is therefore intended to provide realistic spatially correlated material uncertainty while enabling a consistent comparison among the competing reliability-estimation methods.

\begin{figure*}[tb]
\centering
\includegraphics[width=0.7\textwidth]
{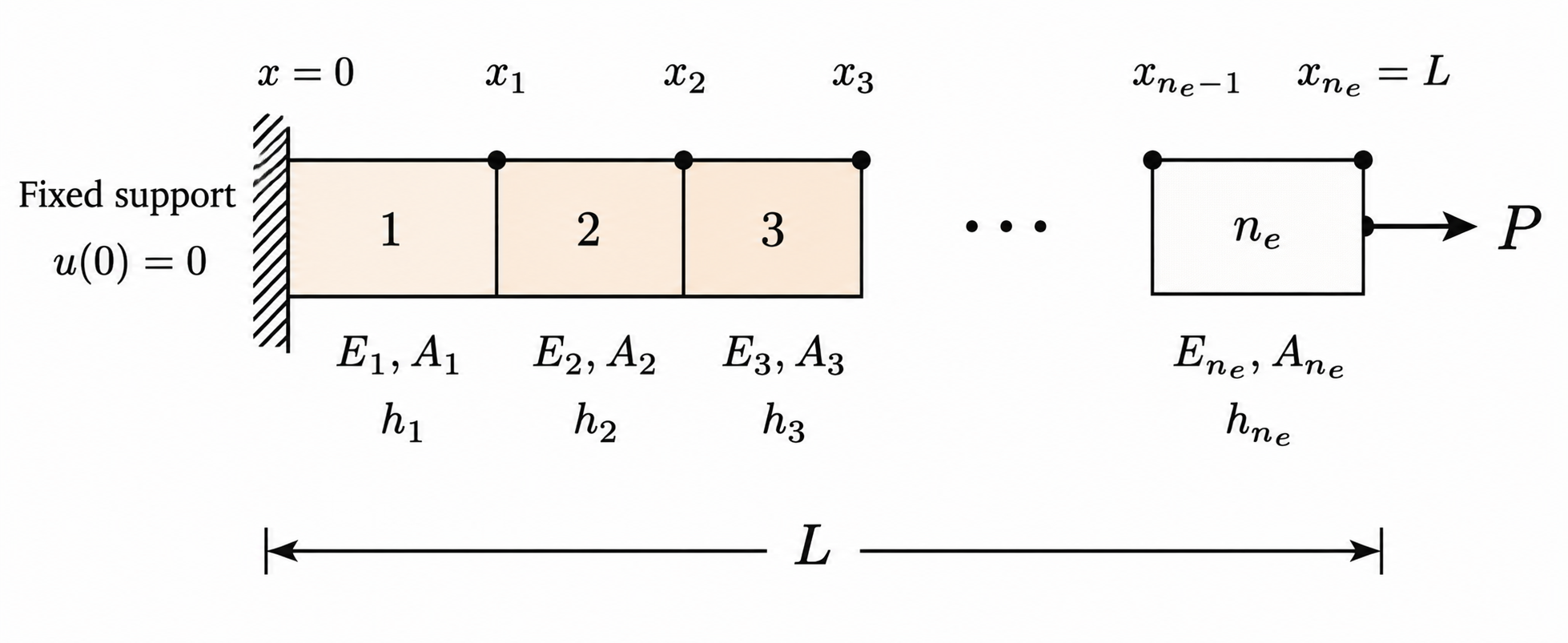}
\caption{Geometry and boundary conditions of the 1D bar. }
\label{fig:1dbar}
\end{figure*}

\subsection{Example 1: Stochastic One-Dimensional Bar}

The first numerical example considers a stochastic one-dimensional axial bar shown in \fref{fig:1dbar} and serves as a controlled benchmark for assessing the proposed Bayesian IQAE framework for rare-event failure-probability estimation. The bar is discretized into \(n_e=20\) finite elements. Uncertainty is introduced through element-wise variations in both the Young's modulus and the cross-sectional area. For each stochastic realization, every element \(e\) is assigned a Young's modulus \(E_e\) and a cross-sectional area \(A_e\) sampled independently from prescribed discrete sets. Specifically, \(E_e\) is drawn from 15 uniformly spaced values in the interval \([0.5,8.0]\), while \(A_e\) is drawn from 15 uniformly spaced values in \([0.5,2.0]\). The samples are generated independently across elements and independently between \(E\) and \(A\), so that no spatial correlation or cross-correlation is imposed in this benchmark. For a given realization, the finite-element stiffness matrix is assembled using the standard axial-bar relation
\begin{equation}
k^{(e)} = \frac{E_e A_e}{h_e}
\begin{bmatrix}
1 & -1\\
-1 & 1
\end{bmatrix},
\label{eq:bar1d_elem_stiffness}
\end{equation}
where \(h_e\) denotes the length of element \(e\).
A fixed stochastic database containing
$N_s = 2^{17}=131{,}072$
realizations is generated prior to the reliability analysis. The realizations are not equally weighted. Instead, each realization \(\boldsymbol{\xi}_i\) is assigned a probability \(p_i\) obtained by normalizing a set of positive random weights,
\begin{equation}
p_i = \frac{\tilde p_i}{\sum_{j=1}^{N_s} \tilde p_j},
\qquad i=1,\dots,N_s,
\label{eq:bar1d_probabilities}
\end{equation}
so that \(\sum_{i=1}^{N_s} p_i = 1\).

For each realization, the finite-element equilibrium equations are solved and the structural response quantity in Eq.~\eqref{eq:framework_limit_state} is taken as the displacement at the loaded end of the bar,
\begin{equation}
Q(\boldsymbol{\xi}) = u(\boldsymbol{\xi}),
\label{eq:bar1d_qoi}
\end{equation}
where \(u(\boldsymbol{\xi})\) denotes the tip displacement for realization \(\boldsymbol{\xi}\). The allowable threshold in Eq.~\eqref{eq:framework_limit_state} is therefore specified in terms of an admissible displacement \(u_{\rm allow}\), and the corresponding limit-state function becomes
\begin{equation}
g(\boldsymbol{\xi}) = u_{\rm allow} - u(\boldsymbol{\xi}).
\label{eq:bar1d_limit_state}
\end{equation}
A realization is classified as failed whenever \(g(\boldsymbol{\xi})<0\), or equivalently whenever the tip displacement exceeds the allowable value. The failure probability is then evaluated according to Eq.~\eqref{eq:framework_pf}. For each realization \(\boldsymbol{\xi}_i\), the displacement response \(u_i\) and the corresponding binary failure indicator \(I_i\) are stored in the stochastic database. The allowable threshold \(u_{\rm allow}\) is selected from the empirical displacement distribution to generate rare-event failure scenarios with prescribed reference failure probabilities.

This benchmark is designed to isolate the statistical behavior of Bayesian IQAE independently of present-day quantum-hardware limitations. 
Measurement outcomes are generated directly from the Bernoulli distribution associated with the amplified probability defined in Eq.~\eqref{eq:amplified_probability}.
The reported oracle budget, therefore, represents the total number of oracle evaluations consumed by the IQAE procedure, while hardware-dependent effects such as gate errors, decoherence, measurement noise, and circuit-compilation overhead are excluded.

The numerical investigation focuses on several aspects of the proposed framework. First, Bayesian IQAE is compared with classical Monte Carlo simulation under identical oracle-evaluation budgets in order to assess estimation accuracy and convergence behavior. Second, the evolution of the posterior distribution is examined as additional oracle resources become available, providing insight into the uncertainty-reduction mechanism described in Section~\ref{sec:sequential_update}. Third, the effect of event rarity on posterior uncertainty is investigated across multiple failure-probability levels. Furthermore, the internal dynamics of a representative Bayesian IQAE run are analyzed to illustrate the interaction between amplitude amplification, posterior updating, and uncertainty quantification during the estimation process.

\begin{table}[tb]
\centering
\caption{Monte Carlo sampling statistics for the rare-event case
($p_f \approx 1.5\times10^{-5}$) based on 30 independent runs.
For each run, realization indices are sampled from the discrete probability distribution
$\{p_i\}_{i=1}^{N_s}$.
The table reports the number of sampled failed realizations.
At small budgets, the expected number of sampled failures is much smaller than one,
so most Monte Carlo runs observe no failures.}
\label{tab:mc_failure_statistics}

\begin{tabular}{cccc}
\hline
Oracle Budget & Mean Failed samples & Median Failed Samples & Fraction of Runs with no Failures \\
\hline
4,000   & 0.033 & 0 & 96.7\% \\
8,000   & 0.100 & 0 & 90.0\% \\
16,000  & 0.367 & 0 & 70.0\% \\
32,000  & 0.267 & 0 & 73.3\% \\
64,000  & 1.167 & 1 & 30.0\% \\
128,000 & 1.700 & 2 & 23.3\% \\
256,000 & 3.667 & 4 & 0.0\% \\
\hline
\end{tabular}

\end{table}

\begin{figure}[tb]
\centering
\begin{subfigure}[t]{0.48\textwidth}
\centering
\includegraphics[width=\textwidth]{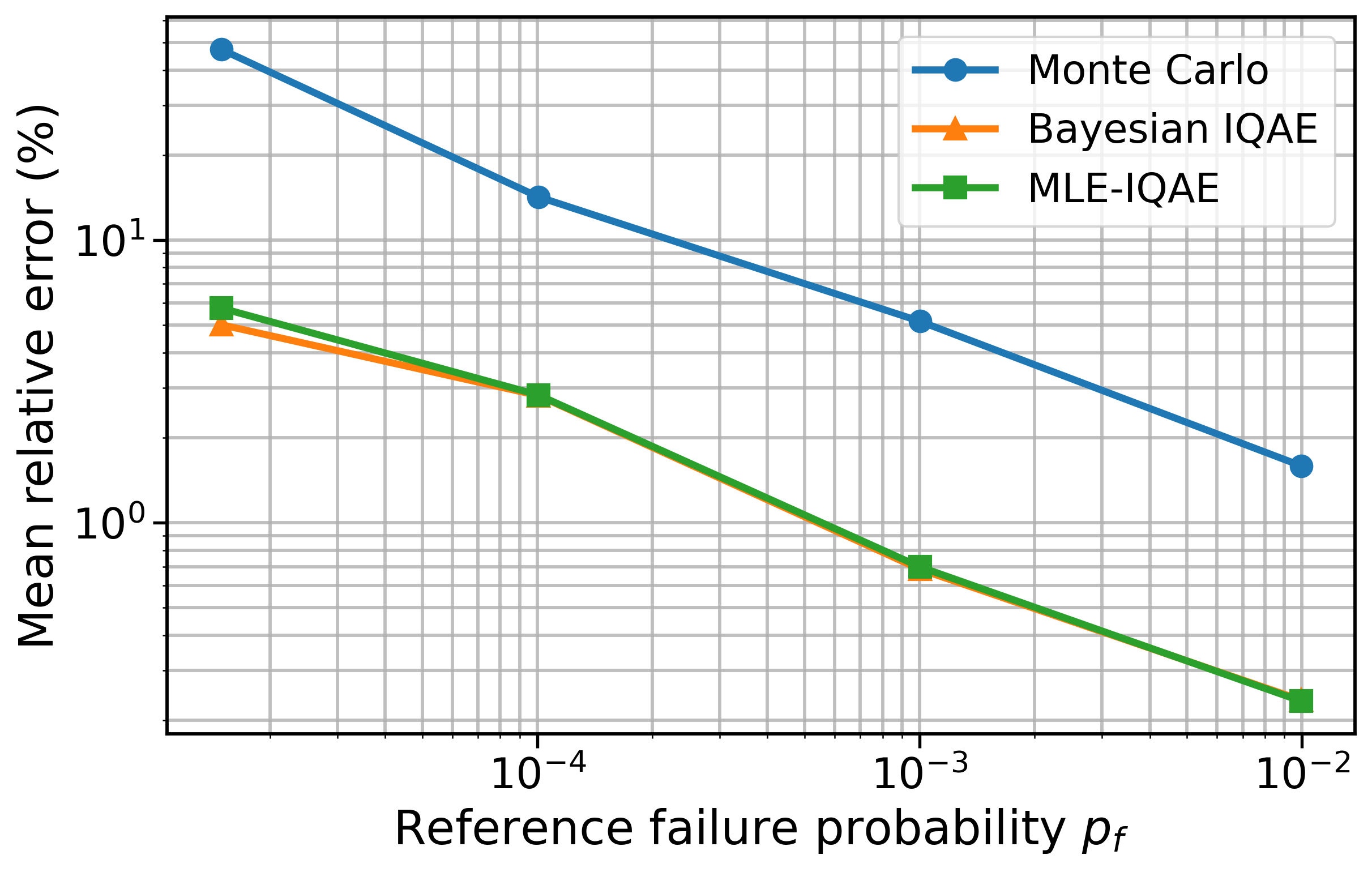}
\caption{Mean relative error as a function of failure probability at a fixed oracle budget of $B=2.56\times10^5$. }
\end{subfigure}
\hfill
\begin{subfigure}[t]{0.48\textwidth}
\centering
\includegraphics[width=\textwidth]{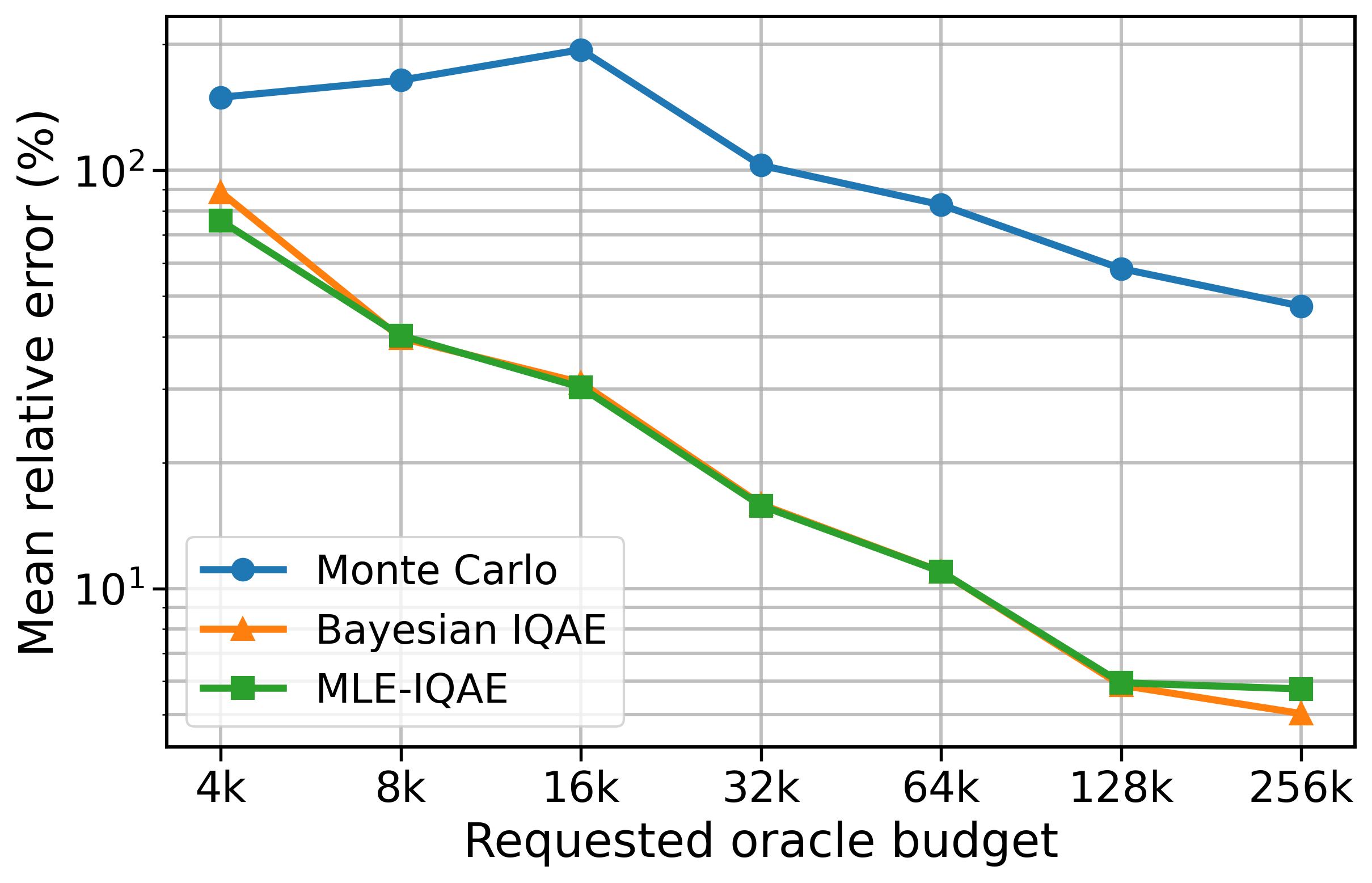}
\caption{Mean relative error as a function of oracle budget for the rare-event case $p_f\approx1.5\times10^{-5}$.}
\end{subfigure}

\vspace{0.25cm}

\begin{subfigure}{0.5\textwidth}
\centering
\includegraphics[width=\textwidth]{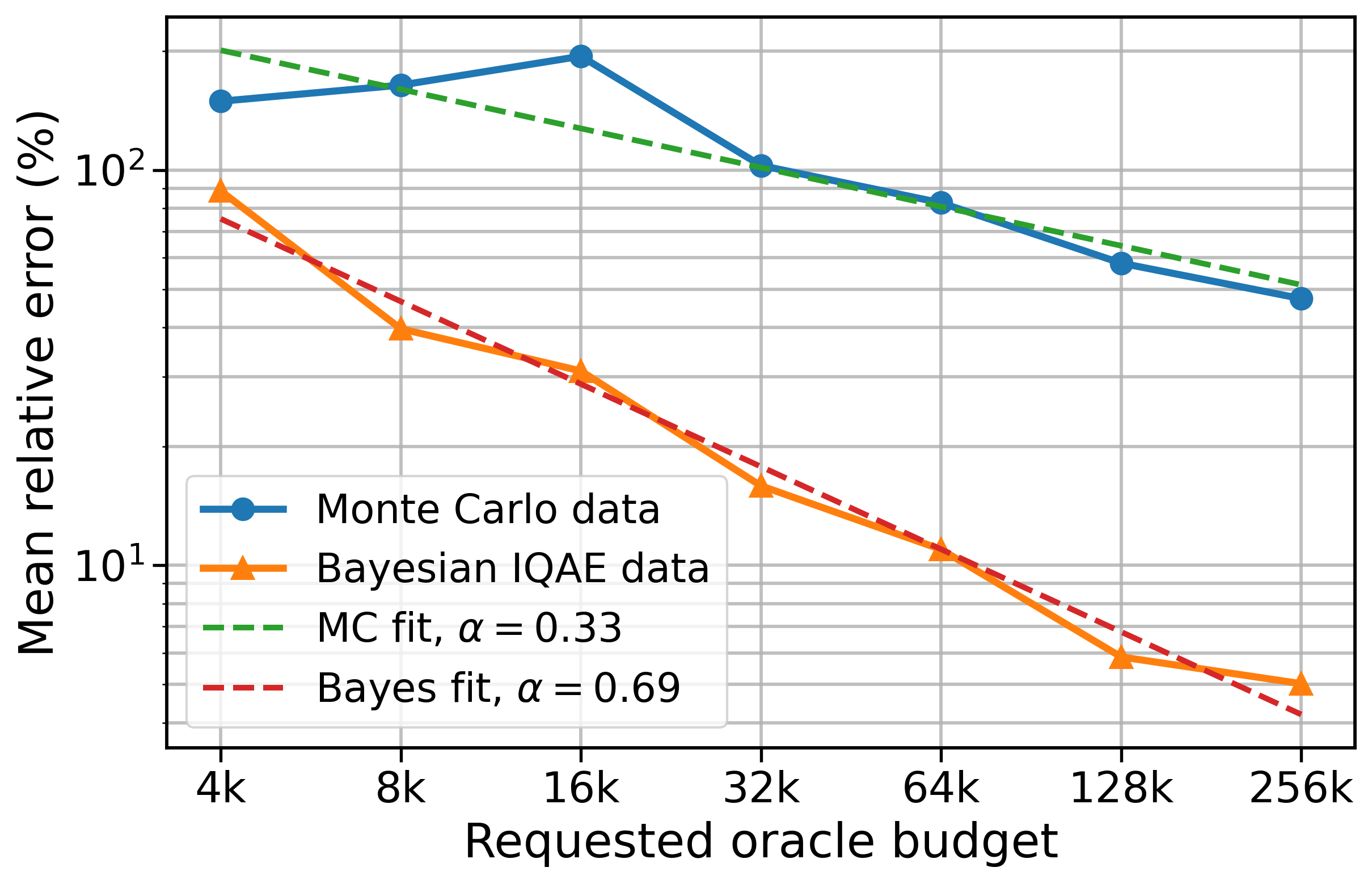}
\caption{Power-law scaling of estimation error with oracle budget for the rare-event case $p_f\approx1.5\times10^{-5}$.}
\end{subfigure}
\caption{Comparison of IQAE and classical Monte Carlo for failure-probability estimation. }
\label{fig:mc_comparison}
\end{figure}

\subsubsection{Sampling Characteristics of Rare Events}
\label{sec:rare_event_sampling}

The rarity of the failure event has important consequences for classical Monte Carlo estimation. For the case considered in this example, the failure probability is
\(p_f \approx 1.5\times10^{-5}\).
Because the realizations in this benchmark are assigned nonuniform probabilities, the failure probability is computed according to the weighted discrete expression in Eq.~\eqref{eq:pf_discrete}, rather than as a simple fraction of failed realizations in the stochastic database. The example, therefore, represents an extremely rare-event reliability problem in which failure carries only a very small probability mass under the discrete stochastic distribution. Such situations are particularly challenging for classical Monte Carlo simulation because large sample counts may be required before any failures are observed.

In the Monte Carlo simulations, realization indices are sampled from the discrete distribution defined by the probabilities \(p_i\). Consequently, each draw corresponds to a Bernoulli trial with success probability \(p_f\), and the expected number of observed failures in a simulation containing \(B\) samples is
\begin{equation}
E[N_f]=Bp_f,
\label{eq:expected_failures}
\end{equation}
where \(N_f\) denotes the number of observed failures. For small sampling budgets, the expected number of failures is substantially less than one, implying that many independent simulations will observe no failures at all. For example, at a budget of \(B=4000\) samples,
\begin{equation}
E[N_f]=4000\times1.5\times10^{-5}\approx 0.06,
\label{eq:expected_failures_4000}
\end{equation}
indicating that the average simulation is expected to observe considerably fewer than a single failure event.

This behavior is quantified in Table~\ref{tab:mc_failure_statistics}, which summarizes the results of 30 independent Monte Carlo simulations for several sampling budgets. At a budget of 4000 samples, the average number of observed failures is only 0.033, and approximately \(96.7\%\) of all runs fail to observe a single failure event. Even when the budget is increased to 16,000 samples, the median number of failures remains zero and \(70.0\%\) of the simulations still observe no failures. Consequently, the majority of Monte Carlo estimates at these budgets are identically zero despite the existence of a nonzero failure probability.

Although the situation improves as the sampling budget increases, failure observations remain sparse throughout much of the investigated range. At a budget of 64,000 samples, the expected number of observed failures is still less than one, and a substantial fraction of the simulations continue to observe no failures. Only at the largest budget considered, \(B=256{,}000\), do all runs observe at least one failure event, with an average of 3.667 failures per simulation. Even then, the estimate is based on only a handful of observed failures, leading to substantial statistical variability.

The results in Table~\ref{tab:mc_failure_statistics} illustrate a fundamental limitation of direct Monte Carlo sampling for rare-event reliability analysis. When failure events are observed only rarely within a simulation, the estimator receives very limited information about the failure probability, resulting in slow convergence and large sampling variability. This challenge motivates the use of amplitude-estimation-based approaches, which seek to extract more information from a fixed computational budget by transforming the underlying rare-event probability into a more readily observable quantity.

\subsubsection{Accuracy and Convergence Comparison}
\label{sec:comparison_mc}

The previous subsection demonstrated that direct Monte Carlo sampling encounters significant difficulties in the rare-event regime because only a small number of failures are observed even at relatively large oracle budgets. The natural question is whether amplitude-estimation-based approaches can utilize the same oracle budget more effectively. To address this question, the performance of Bayesian IQAE, maximum-likelihood IQAE (MLE-IQAE), and classical Monte Carlo is compared using identical oracle budgets.

The estimation accuracy is quantified using the relative error
\begin{equation}
\varepsilon_r
=
\frac{
\left|
\widehat{a}-a
\right|
}
{a},
\label{eq:relative_error}
\end{equation}
where \(a=p_f\) is the reference failure probability and \(\widehat{a}\) denotes the corresponding estimate obtained from Monte Carlo, MLE-IQAE, or Bayesian IQAE.
Figure~\ref{fig:mc_comparison}(a) presents the mean relative estimation error for several failure-probability levels using a fixed oracle budget of $2.56\times10^5$. The considered probabilities span approximately three orders of magnitude, ranging from $O(10^{-5})$ to $O(10^{-2})$. For all probability levels investigated, both Bayesian IQAE and MLE-IQAE produce substantially smaller estimation errors than classical Monte Carlo. The advantage becomes increasingly pronounced as the event becomes rarer. For the smallest probability considered, the estimation error obtained using amplitude estimation is nearly an order of magnitude smaller than that of direct Monte Carlo sampling.

An important observation from Figure~\ref{fig:mc_comparison}(a) is the close agreement between Bayesian IQAE and MLE-IQAE. Across the entire range of failure probabilities considered, both methods exhibit nearly identical estimation accuracy. This result indicates that the Bayesian formulation does not sacrifice accuracy in exchange for uncertainty quantification. Instead, it retains essentially the same estimation performance as maximum-likelihood IQAE while simultaneously providing posterior distributions, credible intervals, and uncertainty-aware stopping criteria.

The influence of oracle budget is examined in Figure~\ref{fig:mc_comparison}(b) for the representative rare-event case with $p_f\approx1.5\times10^{-5}$. As expected, the estimation error decreases for all methods as additional oracle evaluations become available. However, both Bayesian IQAE and MLE-IQAE consistently outperform classical Monte Carlo throughout the entire budget range investigated. The performance gap becomes increasingly apparent as the budget increases, demonstrating that amplitude-estimation-based methods convert additional oracle evaluations into estimation accuracy more efficiently than direct sampling.

The close agreement between Bayesian IQAE and MLE-IQAE is again evident in Figure~\ref{fig:mc_comparison}(b). Although small differences are observed at individual budgets due to finite-shot variability, both methods exhibit nearly identical convergence behavior. This observation is expected because both approaches are ultimately extracting information from the same sequence of amplified measurement outcomes. The primary distinction is that MLE-IQAE reports a single best-fit estimate, whereas Bayesian IQAE maintains and updates an entire posterior distribution over the unknown amplitude.

To further quantify the convergence behavior, the observed errors were fitted using the empirical power-law model
\begin{equation}
\varepsilon_r(B)=CB^{-\alpha},
\label{eq:error_scaling}
\end{equation}
where $B$ denotes the oracle budget, $C$ is a fitting constant, and $\alpha$ characterizes the convergence rate. The resulting fits are shown in Figure~\ref{fig:mc_comparison}(c). For the representative rare-event case, the fitted exponents are approximately
\begin{equation}
\alpha_{\mathrm{MC}} \approx 0.33,
\qquad
\alpha_{\mathrm{MLE}} \approx 0.64,
\qquad
\alpha_{\mathrm{Bayes}} \approx 0.69.
\label{eq:error_scaling_exponents}
\end{equation}
Both amplitude-estimation approaches exhibit substantially faster error reduction than classical Monte Carlo. The slightly larger fitted exponent observed for Bayesian IQAE is empirical and may reflect finite-budget regularization and posterior averaging rather than a distinct asymptotic complexity improvement relative to MLE-IQAE.

\subsubsection{Evolution of a Representative Bayesian IQAE Run}
\label{sec:iqae_evolution}

While Figure~\ref{fig:mc_comparison} demonstrates the superior estimation accuracy of Bayesian IQAE, it does not reveal how the algorithm progressively acquires information about the unknown failure probability. To illustrate the internal behavior of the method, Table~\ref{tab:iqae_evolution_256k} summarizes a representative Bayesian IQAE run for
the rare-event case with a requested oracle budget of $2.56\times10^5$.
\begin{table}[bt]
\centering
\caption{Evolution of a representative Bayesian IQAE run for the rare-event case
with a requested oracle budget of $2.56\times10^5$. The amplified success
probability becomes measurable as the Grover depth increases, while the Bayesian
credible interval for the failure probability continues to contract.}
\label{tab:iqae_evolution_256k}
\begin{tabular}{cccccccc}
\hline
$t$ & $k$ & Shots & Successes & $\hat p_k$ &
Oracle calls & $\theta_{\rm width}$ & $CI_{95}$ \\
\hline
1  & 0  & 533 & 0  & 0.0000 & 533    & 0.088569 & $4.47\times10^{-3}$ \\
2  & 5  & 164 & 0  & 0.0000 & 2337   & 0.016711 & $1.22\times10^{-4}$ \\
3  & 10 & 261 & 0  & 0.0000 & 7818   & 0.007432 & $1.82\times10^{-5}$ \\
4  & 15 & 385 & 7  & 0.0182 & 19753  & 0.005274 & $2.08\times10^{-5}$ \\
5  & 20 & 434 & 8  & 0.0184 & 37547  & 0.003458 & $1.24\times10^{-5}$ \\
6  & 25 & 532 & 16 & 0.0301 & 64679  & 0.002906 & $8.42\times10^{-6}$ \\
7  & 30 & 448 & 29 & 0.0647 & 92007  & 0.002043 & $6.70\times10^{-6}$ \\
8  & 33 & 407 & 33 & 0.0811 & 119276 & 0.001922 & $6.30\times10^{-6}$ \\
9  & 36 & 374 & 26 & 0.0695 & 146578 & 0.001922 & $5.25\times10^{-6}$ \\
10 & 39 & 346 & 30 & 0.0867 & 173912 & 0.001922 & $4.91\times10^{-6}$ \\
11 & 42 & 241 & 28 & 0.1162 & 194397 & 0.001922 & $4.23\times10^{-6}$ \\
\hline
\end{tabular}
\end{table}
Several important observations emerge from the table. During the initial iterations, the algorithm employs relatively small Grover depths and observes no successful measurements. For example, the first three iterations produce zero successes despite consuming more than 7,800 oracle evaluations. Such outcomes are expected because the underlying failure probability is extremely small. Nevertheless, the absence of observed failures is itself informative. Even without a single positive observation, the Bayesian update process rapidly eliminates large portions of the admissible parameter space, reducing the interval width in $\theta$ from $8.86\times10^{-2}$ to $7.43\times10^{-3}$ within the first three iterations. This behavior highlights an important advantage of the Bayesian formulation; information is extracted not only from successful measurements but also from the repeated absence of failures.

As the Grover depth increases, amplitude amplification transforms the underlying rare event into a directly measurable quantity. Beginning at iteration $t=4$, nonzero success counts appear, and the amplified measurement probabilities increase steadily with increasing amplification level. By iteration $t=11$, the measured success probability has reached approximately $11.6\%$, even though the underlying failure probability remains on the order of $10^{-5}$. Consequently, Bayesian IQAE is able to collect informative measurement data in a regime where direct Monte Carlo sampling would still observe only a handful of failures.

\begin{figure*}[tb]
\centering
\begin{subfigure}[t]{0.48\textwidth}
\centering
\includegraphics[width=\textwidth]{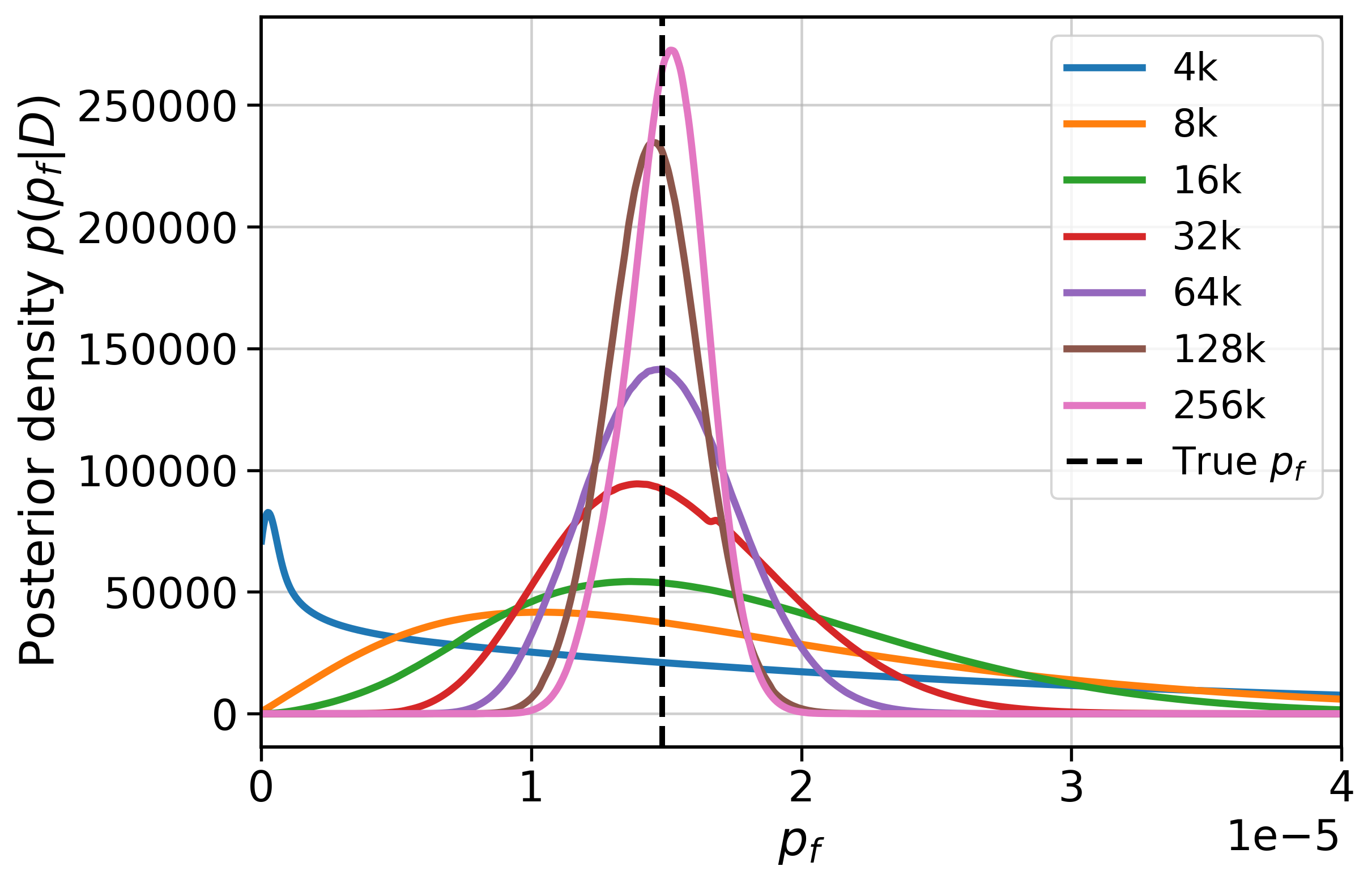}
\caption{Posterior density functions for different oracle budgets. Increasing oracle budget leads to progressively stronger concentration of posterior mass around the reference failure probability.}
\label{fig:posterior_density_budget}
\end{subfigure}
\hfill
\begin{subfigure}[t]{0.48\textwidth}
\centering
\includegraphics[width=\textwidth]{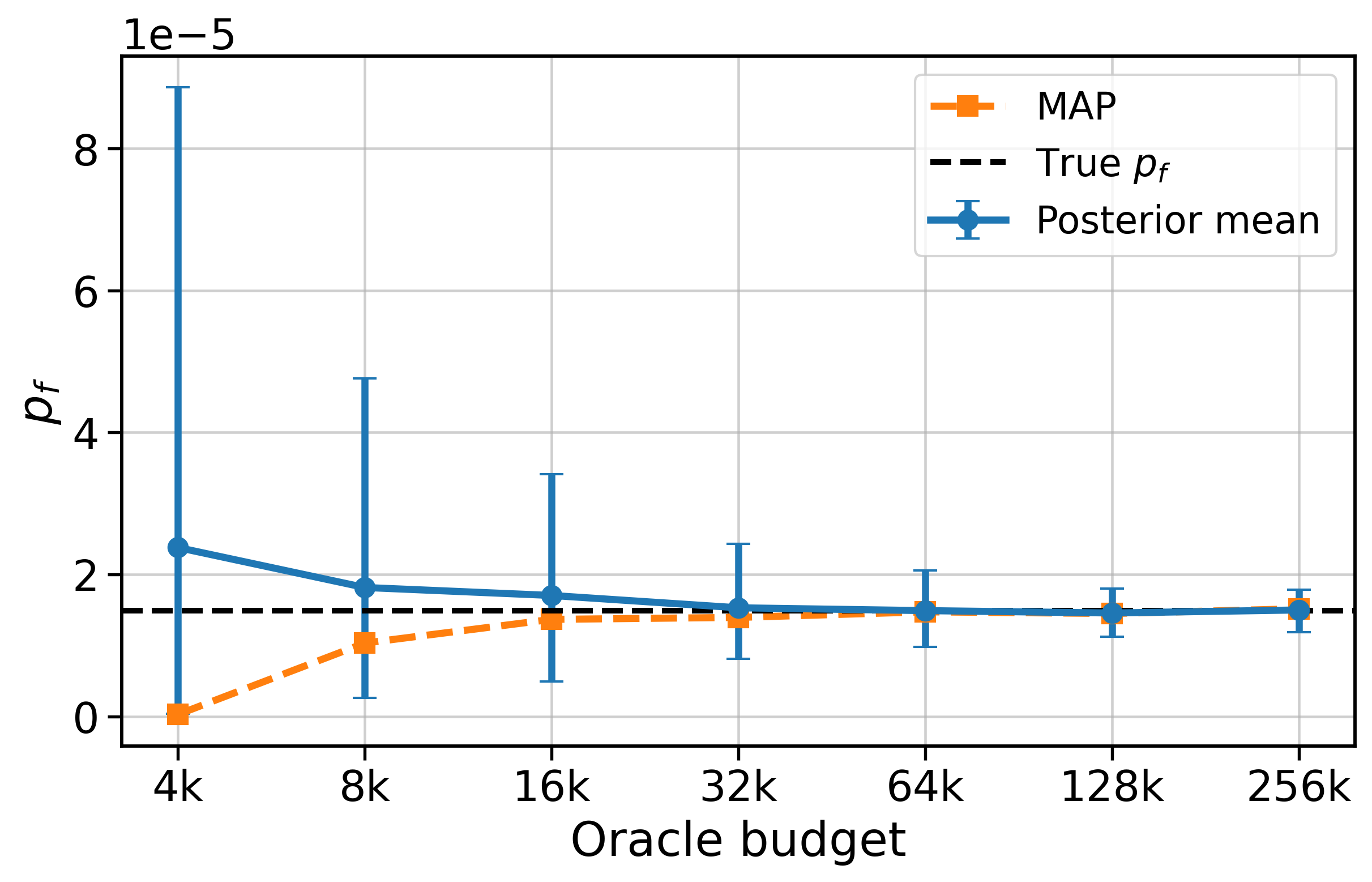}
\caption{Posterior mean, MAP estimate, and associated $95\%$ credible interval as functions of oracle budget.}
\label{fig:posterior_mean_map_ci}
\end{subfigure}

\vspace{0.15in}

\begin{subfigure}[t]{0.48\textwidth}
\centering
\includegraphics[width=\textwidth]{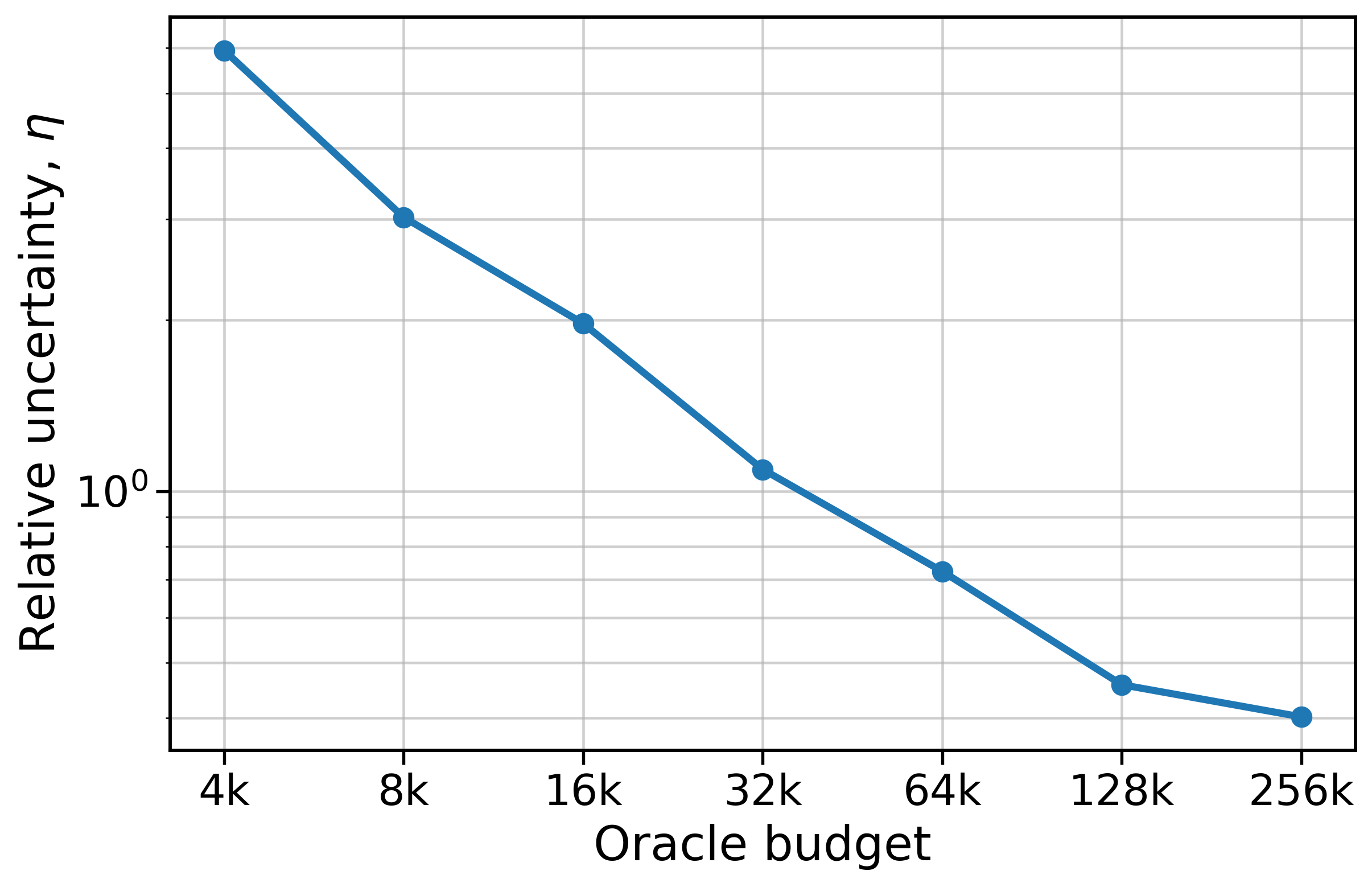}
\caption{Relative uncertainty measure
$\eta=\Delta CI_{95}/p_f$
as a function of oracle budget, showing systematic posterior contraction and uncertainty reduction.}
\label{fig:posterior_ci_width}
\end{subfigure}

\caption{Bayesian posterior contraction with increasing oracle budget for $p_f\approx1.5\times10^{-5}$.}
\label{fig:bayesian_posterior_contraction}
\end{figure*}

The most significant trend in Table~\ref{tab:iqae_evolution_256k} is the evolution of posterior uncertainty. The final column reports the width of the $95\%$ credible interval for the inferred failure probability. Over the course of the run, this interval contracts from $4.47\times10^{-3}$ to $4.23\times10^{-6}$, corresponding to a reduction of more than three orders of magnitude. Such contraction demonstrates how Bayesian IQAE progressively concentrates posterior probability mass around the true failure probability as additional oracle information becomes available.

An especially noteworthy observation occurs during the later stages of the algorithm. Beginning at approximately iteration $t=8$, the width of the admissible interval for the amplitude angle $\theta$ remains essentially unchanged at
$\theta_{\rm width}\approx1.92\times10^{-3}$. From the perspective of interval-based IQAE methods, this behavior could be interpreted as an indication that convergence has largely been achieved. The Bayesian posterior, however, reveals that additional information continues to be extracted from the quantum measurements.
Although the admissible region in $\theta$ changes very little after iteration $t=8$, the posterior distribution continues to concentrate within that region. Consequently, the width of the $95\%$ credible interval, $w_{95}$, decreases from $6.30\times10^{-6}$ at iteration $t=8$ to $4.23\times10^{-6}$ at iteration $t=11$, corresponding to an additional reduction of approximately $33\%$. This behavior indicates that the dominant branch ambiguity has already been resolved and that subsequent measurements are primarily refining the distribution of probability mass within the surviving region of parameter space.
This distinction highlights an important advantage of the Bayesian formulation. Whereas interval-based approaches monitor only the size of the admissible parameter region, Bayesian IQAE tracks the complete posterior distribution and therefore remains sensitive to continued reductions in uncertainty even after the interval estimate has largely stabilized. The posterior thus provides a more informative measure of convergence than interval width alone.

Overall, Table~\ref{tab:iqae_evolution_256k} provides insight into the mechanism responsible for the improved performance observed in Figure~\ref{fig:mc_comparison}. Through the combination of amplitude amplification and sequential Bayesian inference, the algorithm progressively converts information from quantum measurements into reductions in posterior uncertainty, ultimately producing accurate estimates of extremely small failure probabilities.
These results have important implications for rare-event structural reliability analysis. For the representative case with $p_f\approx1.5\times10^{-5}$, classical Monte Carlo requires hundreds of thousands of samples before failure events are observed consistently. Even at such budgets, the estimate is based on only a handful of failures and therefore remains highly variable. In contrast, Bayesian IQAE remains effective because amplitude amplification transforms the underlying rare-event probability into a more informative measurement process, while Bayesian updating systematically combines information across multiple Grover depths. Consequently, the proposed framework is particularly attractive for structural certification, extreme-value analysis, and tail-risk estimation problems in which rare events play a dominant role in design and decision making.

\subsubsection{Posterior Contraction with Increasing Oracle Budget}
\label{sec:bayesian_posterior_contraction}

An important feature of Bayesian IQAE is that it produces a posterior distribution for the unknown failure probability rather than a single point estimate. The evolution of this posterior provides direct insight into how uncertainty is reduced as additional quantum measurements become available.
To investigate the evolution of posterior uncertainty, Bayesian IQAE simulations were performed for the representative rare-event case with $p_f\approx1.5\times10^{-5}$ using oracle budgets ranging from $4\times10^3$ to $1.28\times10^5$ oracle evaluations. For each budget level, 30 independent runs were conducted using different random seeds. Figure~\ref{fig:bayesian_posterior_contraction} summarizes the resulting posterior distributions and associated uncertainty measures.

Figure~\ref{fig:bayesian_posterior_contraction}(a) presents the posterior density functions corresponding to different oracle budgets. At small budgets, the posterior distribution is relatively broad, reflecting substantial uncertainty regarding the location of the failure probability. As additional oracle evaluations become available, the posterior progressively concentrates around the reference solution. This behavior is a direct manifestation of Bayesian posterior contraction and demonstrates that the information collected through successive IQAE measurements is effectively translated into increasingly precise probabilistic inferences.

The increasing peak heights observed in Figure~\ref{fig:bayesian_posterior_contraction}(a) should not be interpreted as an increase in probability mass. Since every posterior density satisfies
\begin{equation}
\int p(p_f|D),dp_f = 1,
\label{eq:posterior_normalization}
\end{equation}
the total probability mass remains constant. Instead, the higher peaks arise because the same probability mass becomes concentrated within a progressively smaller region of parameter space. Consequently, narrower and taller posterior distributions correspond directly to reduced uncertainty regarding the unknown failure probability.

Additional insight is provided by Figure~\ref{fig:bayesian_posterior_contraction}(b), which reports the posterior mean, MAP estimate, and corresponding $95\%$ credible interval as functions of oracle budget. At small budgets, the posterior mean and MAP estimate exhibit noticeable variability and the credible intervals remain relatively wide. As the oracle budget increases, both estimators converge toward the reference solution while the credible intervals contract substantially. For the largest budgets considered, the posterior mean, MAP estimate, and reference solution become nearly indistinguishable on the scale of the figure.

The reduction in posterior uncertainty is quantified more directly in Figure~\ref{fig:bayesian_posterior_contraction}(c), which reports the relative uncertainty measure
$\eta=\Delta CI_{95}/p_f$
as a function of oracle budget. A systematic decrease in $\eta$ is observed throughout the entire budget range, indicating continuous posterior contraction as additional oracle evaluations are incorporated. For the smallest budget considered, the width of the $95\%$ credible interval exceeds the magnitude of the failure probability itself, reflecting substantial uncertainty in the estimate. As the oracle budget increases, $\eta$ decreases steadily and falls below unity, indicating that the uncertainty becomes smaller than the target failure probability. Relative to the smallest-budget case, the largest-budget case exhibits approximately an order-of-magnitude reduction in normalized uncertainty. This trend demonstrates that additional oracle evaluations not only improve estimation accuracy but also increase confidence in the inferred failure probability.

\begin{figure}[t]
\centering
\includegraphics[width=\textwidth]{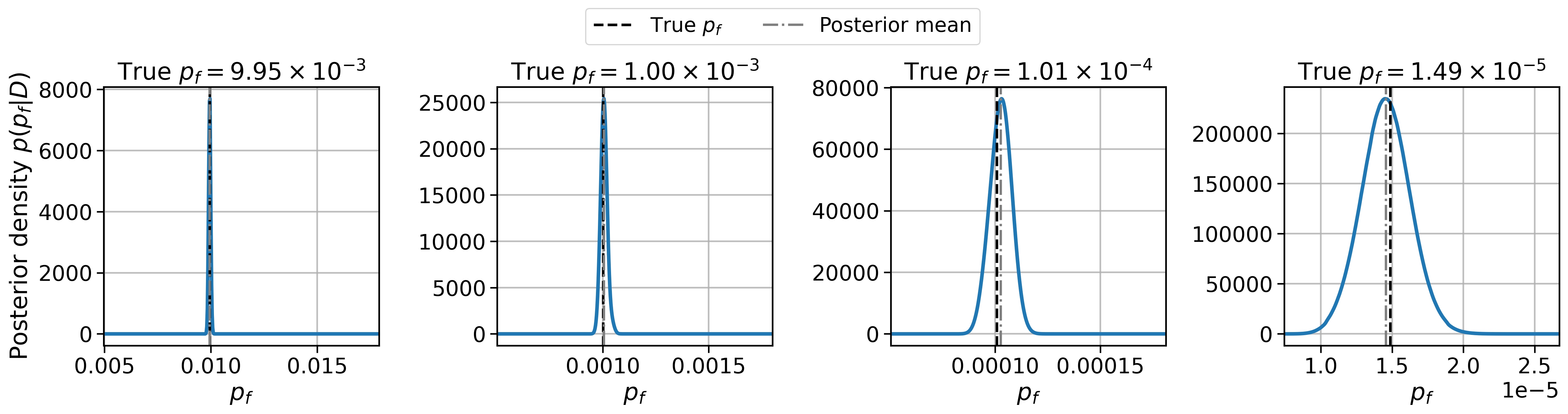}
\caption{Posterior probability density functions obtained using Bayesian IQAE for four failure-probability levels at an oracle budget of $1.28\times10^5$. The dashed black line denotes the reference failure probability and the dash-dotted gray line denotes the posterior mean. The posterior distribution broadens as the failure event becomes increasingly rare.}
\label{fig:posterior_rarity_physical}
\end{figure}

\subsubsection{Effect of Event Rarity on Posterior Uncertainty}
\label{sec:event_rarity}

The previous subsection demonstrated that, for a fixed reliability problem, the Bayesian IQAE posterior contracts systematically as additional oracle evaluations become available. An equally important question is how the posterior behaves when the underlying event itself becomes increasingly rare. In practical reliability analysis, probabilities of interest often span several orders of magnitude, ranging from moderately rare events to extremely small tail probabilities. Consequently, it is important to assess whether Bayesian IQAE maintains both accuracy and meaningful uncertainty quantification as the estimation problem becomes progressively more challenging.

To investigate this effect, four reliability scenarios were considered corresponding to reference failure probabilities of approximately $3\times10^{-2}$, $10^{-3}$, $10^{-4}$, and $10^{-5}$. These probabilities were obtained by varying the failure threshold in the stochastic bar problem while maintaining the same underlying stochastic model. For each case, Bayesian IQAE was executed using a fixed oracle budget of $1.28\times10^5$ evaluations.

Figure~\ref{fig:posterior_rarity_physical} presents the resulting posterior probability density functions on their physical probability scales. The dashed black line denotes the reference failure probability, while the dash-dotted gray line denotes the posterior mean. For all probability levels considered, the posterior distributions remain centered near the reference solution, indicating that Bayesian IQAE provides accurate estimates over several orders of magnitude in failure probability.

\begin{figure}[t]
\centering

\begin{subfigure}[t]{0.48\textwidth}
\centering
\includegraphics[width=\textwidth]{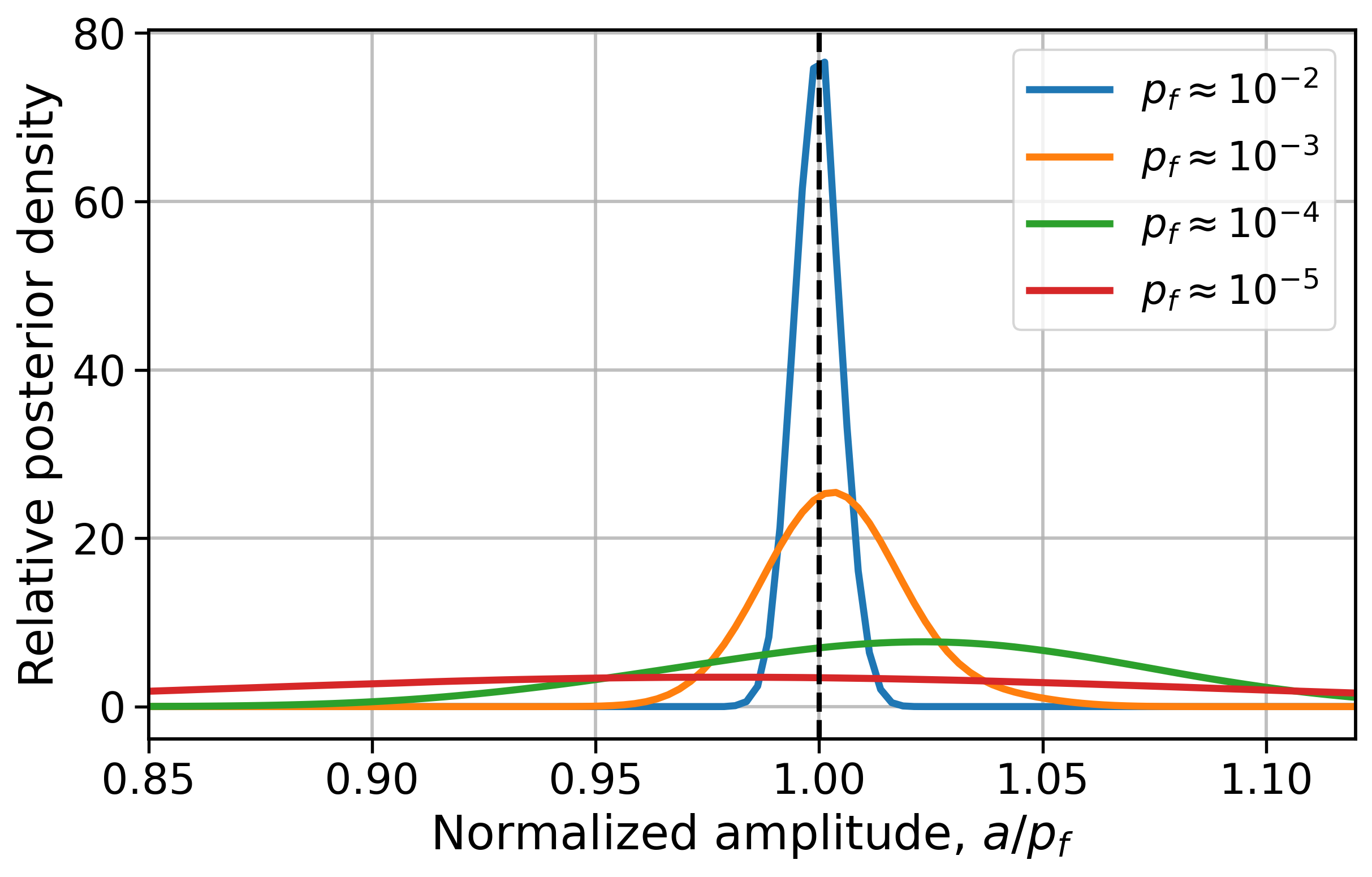}
\caption{Normalized posterior distributions for different failure-probability levels.}
\label{fig:posterior_rarity_relative}
\end{subfigure}
\hfill
\begin{subfigure}[t]{0.48\textwidth}
\centering
\includegraphics[width=\textwidth]{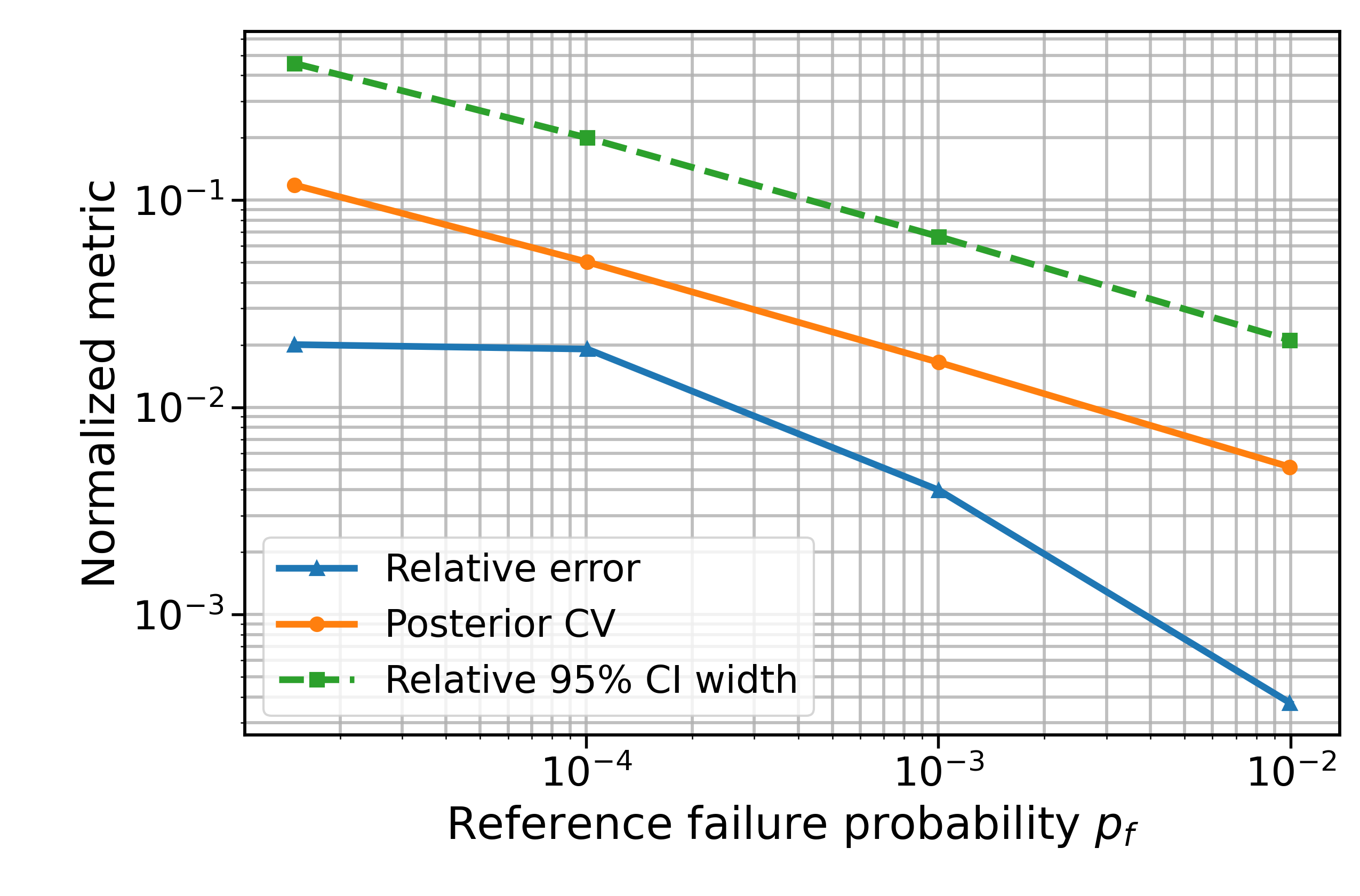}
\caption{Summary measures of estimation accuracy and posterior uncertainty.}
\label{fig:rarity_metrics}
\end{subfigure}

\caption{Effect of event rarity on Bayesian posterior inference at an oracle budget of $2.56\times10^5$.}
\label{fig:event_rarity}
\end{figure}

A clear increase in posterior spread is nevertheless observed as the event becomes rarer. The posterior corresponding to $p_f\approx3\times10^{-2}$ is sharply concentrated around the reference value, whereas the posterior associated with $p_f\approx 1.5 \times 10^{-5}$ exhibits substantially greater dispersion. This behavior reflects the decreasing amount of information available when estimating increasingly rare events and demonstrates that the Bayesian posterior appropriately adapts to the intrinsic difficulty of the underlying inference problem.

While Figure~\ref{fig:posterior_rarity_physical} presents the posterior distributions on their original physical scale, direct comparison across several orders of magnitude is difficult because each case is associated with a different failure-probability level. To facilitate comparison, the posterior distributions were normalized by the exact failure probability of the corresponding case and are shown in Figure~\ref{fig:event_rarity}(a). The horizontal axis therefore represents the ratio \(\hat{a}_B/p_f\), where \(\hat{a}_B\) denotes the Bayesian posterior estimate of the amplitude and \(p_f\) is the reference failure probability of the corresponding stochastic database. Since the amplitude satisfies \(a=p_f\), the vertical dashed line at \(\hat{a}_B/p_f=1\) indicates perfect agreement with the reference solution, allowing all posterior distributions to be compared on a common relative scale.

The normalized posterior distributions reveal a systematic increase in relative uncertainty as the event becomes rarer. Although all distributions remain centered near the reference solution, their relative spread increases significantly as the failure probability decreases from approximately $10^{-2}$ to $10^{-5}$. This behavior is consistent with the diminishing amount of information available about increasingly rare events and demonstrates that the Bayesian posterior appropriately reflects the statistical difficulty of the estimation problem.

To quantify these trends further, Figure~\ref{fig:event_rarity}(b) reports three complementary measures of estimation accuracy and posterior uncertainty: the relative error of the posterior mean,
$
|\widehat{a}_{\mathrm{B}}-p_f|/p_f,
$
the posterior coefficient of variation $\mathrm{CV}_{\mathrm{B}}(a)$ defined in Eq.~\eqref{eq:posterior_cv}, and the normalized width of the $95\%$ credible interval, $\eta$, defined in Eq.~\eqref{eq:relative_credible_width}. Together, these quantities characterize both the accuracy of the posterior estimate and the uncertainty remaining in the inferred failure probability.
The quantitative metrics confirm the behavior observed in Figure~\ref{fig:event_rarity}(a). As the failure probability decreases by approximately three orders of magnitude, both the posterior coefficient of variation and the normalized credible-interval width increase substantially, indicating progressively larger uncertainty in the inferred probability. Nevertheless, the relative error of the posterior mean remains small throughout the entire range of probabilities considered. Even for the rarest event examined, Bayesian IQAE produces an estimate that remains close to the reference solution while simultaneously providing a realistic assessment of the uncertainty associated with the inference.

Importantly, the increase in posterior uncertainty should not be viewed as a deficiency of the Bayesian approach. Rather, it represents an appropriate probabilistic response to the reduced information content associated with increasingly rare events. Instead of reporting only a point estimate, the Bayesian formulation explicitly quantifies this loss of information through broader posterior distributions and wider credible intervals.


\subsection{Stochastic L-Shaped Bracket}
\label{sec:lbracket}

The second numerical example considers a stochastic two-dimensional L-shaped bracket under plane stress conditions, subjected to uncertainty in its material properties. 
Example 1 focused on the statistical properties of Bayesian IQAE using a synthetic reliability benchmark. 
The domain is discretized using a structured finite-element mesh consisting of 4,800 bilinear quadrilateral elements. 
The left boundary of the domain is fully constrained. Therefore, both displacement components are set to zero along \(x=0\)
\begin{equation}
u_x = 0, \qquad u_y = 0 \qquad \text{on } x=0.
\label{eq:lbracket_bc_fixed}
\end{equation}
A uniform vertical traction is applied on the right boundary segment \(x=1\), \(0\le y\le 0.5\). The applied traction has no horizontal component and acts in the negative \(y\)-direction,
\begin{equation}
\mathbf{t} =
\begin{bmatrix}
0 \\
-5
\end{bmatrix}
\qquad \text{on }  x=1,\; 0\le y\le 0.5.
\label{eq:lbracket_bc_load}
\end{equation}
All remaining boundaries are traction-free. This loading and boundary-condition configuration produces a stress concentration near the interior corner of the L-shaped domain.

The Young's modulus in the L-shaped benchmark is modeled as a spatially varying lognormal random field in order to represent material uncertainty while preserving positive stiffness throughout the domain. Spatial correlation is introduced through a Gaussian covariance kernel~\cite{ghanem1991stochastic,xiu2010numerical} with correlation lengths \(\ell_x=0.35L_x=0.35\) and \(\ell_y=0.35L_y=0.35\), and the lognormal standard deviation is set to \(\sigma=0.35\). The covariance operator is approximated using a Nystr\"om discretization~\cite{bate2008efficient} with \(M=350\) sampling points and subsequently truncated to rank \(r=80\), providing a low-rank representation for generating stochastic realizations of the modulus field.

For each stochastic realization, the finite-element equilibrium equations are solved and the maximum von Mises stress is extracted as the quantity of interest. Failure is defined through the stress-based limit-state function
\begin{equation}
g(\omega)
=
\sigma_{\rm allow}
-
\max_{\mathbf{x}\in\Omega}
\sigma_{\rm vm}(\mathbf{x},\omega),
\label{eq:lbracket_limit_state}
\end{equation}
where $\sigma_{\rm allow}$ denotes the allowable stress threshold and $\sigma_{\rm vm,max}$ is the maximum von Mises stress obtained from the finite-element solution. A realization is considered failed whenever
$
g(\omega)<0.
$
The corresponding failure probability is
$
p_f
=
\mathbb{P}
\!\left[
g(\omega)<0
\right],
$
which is estimated using Bayesian IQAE, MLE-IQAE, and classical Monte Carlo simulation.
\begin{figure*}[tb]
\centering
\includegraphics[width=0.45\textwidth]
{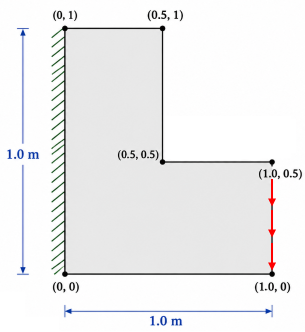}
\caption{Geometry and boundary conditions of the L-shaped domain. }
\label{fig:lbracket}
\end{figure*}

A reference database consisting of $2^{14}=16,384$ stochastic finite-element realizations is generated. Three failure thresholds are selected, producing failure probabilities
\begin{equation}
p_f
\in
\left\{
1.001\times10^{-2},
\;
1.038\times10^{-3},
\;
1.221\times10^{-4}
\right\},
\label{eq:lbracket_pf_levels}
\end{equation}
which span approximately two orders of magnitude in event rarity. These three cases provide a systematic assessment of estimator performance as the underlying failure event becomes increasingly rare.
For each failure probability, Bayesian IQAE, MLE-IQAE, and Monte Carlo simulation are executed using identical oracle budgets ranging from 500 to 20,000 evaluations. All reported statistics are averaged over multiple independent runs.

\subsubsection{Comparison with Monte Carlo and MLE-IQAE}
\label{sec:lbracket_comparison}

\begin{figure*}[t]
\centering
\begin{subfigure}[t]{0.32\textwidth}
\centering
\includegraphics[width=\textwidth]
{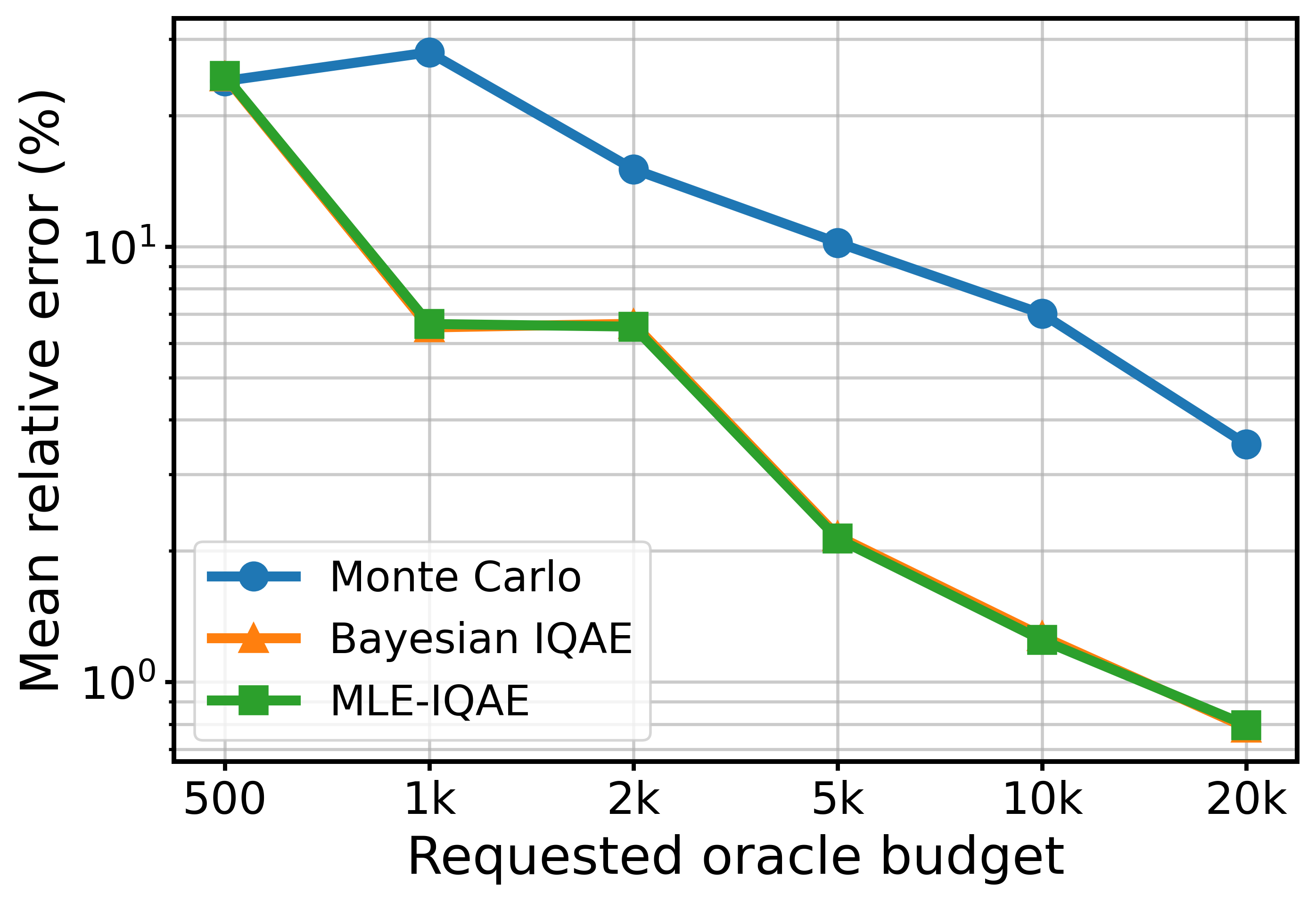}
\caption{$p_f = 1.001\times10^{-2}$}
\label{fig:lbracket_q99_error}
\end{subfigure}
\hfill
\begin{subfigure}[t]{0.32\textwidth}
\centering
\includegraphics[width=\textwidth]
{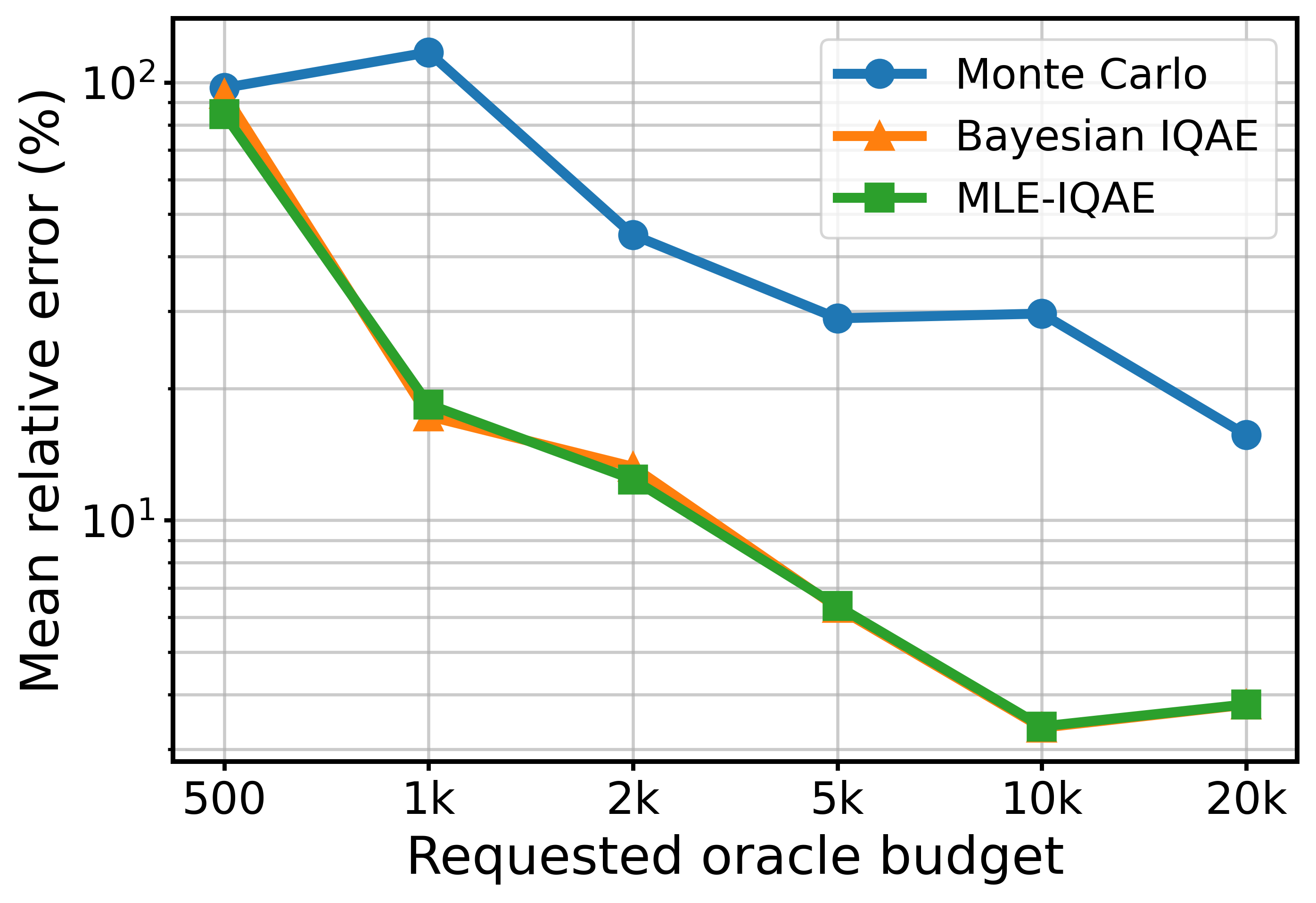}
\caption{$p_f = 1.038\times10^{-3}$}
\label{fig:lbracket_q999_error}
\end{subfigure}
\hfill
\begin{subfigure}[t]{0.32\textwidth}
\centering
\includegraphics[width=\textwidth]
{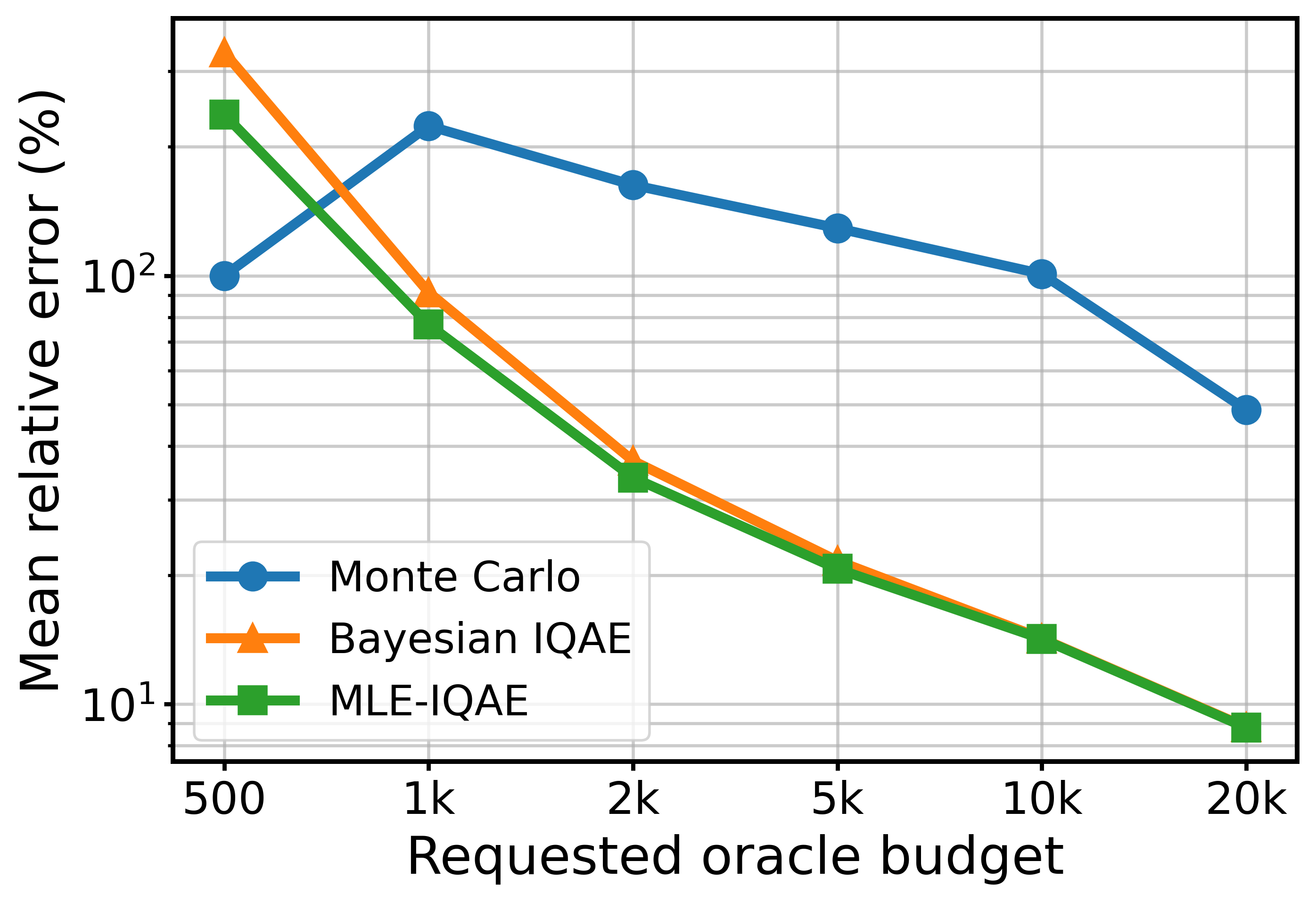}
\caption{$p_f = 1.221\times10^{-4}$}
\label{fig:lbracket_q9999_error}
\end{subfigure}
\caption{
Mean relative error obtained using Monte Carlo simulation, Bayesian IQAE, and MLE-IQAE for three L-bracket failure probabilities. The advantage of the IQAE-based estimators becomes more pronounced as the failure probability decreases. Bayesian IQAE and MLE-IQAE exhibit nearly identical point-estimation accuracy, while both outperform direct Monte Carlo sampling for the same oracle budget.
}
\label{fig:lbracket_error_comparison}
\end{figure*}

Figure~\ref{fig:lbracket_error_comparison} compares the mean relative error obtained using Monte Carlo simulation, Bayesian IQAE, and MLE-IQAE for the three L-bracket failure probabilities. The comparison is performed using the same oracle budgets for all methods.
\begin{table*}[tb]
\centering
\caption{Comparison of Monte Carlo sampling statistics and Bayesian IQAE
accuracy for the three failure probabilities considered.
The Monte Carlo success count denotes the average number of observed
failures over all independent runs.}
\label{tab:lbracket_sampling_statistics}
\begin{tabular}{cccccc}
\hline
$p_f$ &
Budget &
Bayesian Error (\%) &
MC Error (\%) &
MC Success Count &
MC Zero-Failure Rate (\%) \\
\hline

\multirow{6}{*}{$1.001\times10^{-2}$}
& 500   & 24.6 & 24.0 & 5.0   & 0 \\
& 1000  & 6.5  & 28.0 & 10.6  & 0 \\
& 2000  & 6.6  & 15.1 & 17.8  & 0 \\
& 5000  & 2.2  & 10.2 & 47.3  & 0 \\
& 10000 & 1.3  & 7.0  & 99.8  & 0 \\
& 20000 & 0.8  & 3.5  & 200.4 & 0 \\
\hline

\multirow{6}{*}{$1.038\times10^{-3}$}
& 500   & 94.2 & 97.1 & 0.4  & 60 \\
& 1000  & 17.3 & 117.1 & 1.2 & 50 \\
& 2000  & 13.2 & 44.8 & 1.9  & 10 \\
& 5000  & 6.3  & 28.9 & 5.1  & 0 \\
& 10000 & 3.4  & 29.6 & 10.0 & 0 \\
& 20000 & 3.8  & 15.7 & 22.3 & 0 \\
\hline

\multirow{6}{*}{$1.221\times10^{-4}$}
& 500   & 332.5 & 100.0 & 0.0 & 100 \\
& 1000  & 91.5  & 223.8 & 0.2 & 80 \\
& 2000  & 37.0  & 162.9 & 0.3 & 70 \\
& 5000  & 21.6  & 129.2 & 0.3 & 90 \\
& 10000 & 14.2  & 101.0 & 1.1 & 50 \\
& 20000 & 8.8   & 48.7  & 2.5 & 0 \\
\hline
\end{tabular}
\end{table*}
\begin{table*}[tb]
\centering
\caption{Representative Bayesian IQAE evolution for the rare-event case $p_f=1.221\times10^{-4}$ with an oracle budget of approximately $2.0\times10^{4}$. As the amplification depth increases, the amplified success probability becomes measurable, the admissible amplitude interval contracts, and the relative Bayesian credible interval width decreases systematically.}
\label{tab:lbracket_iqae_trace}
\begin{tabular}{ccccccccc}
\hline
$t$ &
$k$ &
Shots &
Successes &
$\hat p_k$ &
Oracle Calls &
$\theta_{\rm width}$ &
$\Delta CI_{95}$ &
$\eta$ \\
\hline
1  & 0  & 281 & 0  & 0.000 &   281  & 0.08857 & $8.47\times10^{-3}$ & 69.4 \\
2  & 5  & 87  & 0  & 0.000 &  1238  & 0.01721 & $2.28\times10^{-4}$ & 1.87 \\
3  & 10 & 81  & 3  & 0.037 &  2939  & 0.01084 & $1.54\times10^{-4}$ & 1.26 \\
4  & 15 & 55  & 8  & 0.145 &  4644  & 0.00914 & $1.35\times10^{-4}$ & 1.11 \\
5  & 20 & 41  & 6  & 0.146 &  6325  & 0.00668 & $9.90\times10^{-5}$ & 0.81 \\
6  & 25 & 33  & 11 & 0.333 &  8008  & 0.00573 & $8.50\times10^{-5}$ & 0.70 \\
7  & 30 & 28  & 10 & 0.357 &  9716  & 0.00454 & $7.10\times10^{-5}$ & 0.58 \\
8  & 33 & 25  & 11 & 0.440 & 11391  & 0.00401 & $6.20\times10^{-5}$ & 0.51 \\
9  & 36 & 23  & 17 & 0.739 & 13070  & 0.00382 & $5.90\times10^{-5}$ & 0.48 \\
10 & 36 & 23  & 12 & 0.522 & 14749  & 0.00348 & $5.30\times10^{-5}$ & 0.43 \\
11 & 36 & 17  & 5  & 0.294 & 15990  & 0.00311 & $4.80\times10^{-5}$ & 0.39 \\
12 & 39 & 12  & 7  & 0.583 & 16938  & 0.00298 & $4.60\times10^{-5}$ & 0.38 \\
13 & 39 & 9   & 5  & 0.556 & 17649  & 0.00291 & $4.40\times10^{-5}$ & 0.36 \\
14 & 40 & 7   & 5  & 0.714 & 18216  & 0.00286 & $4.40\times10^{-5}$ & 0.36 \\
\hline
\end{tabular}
\end{table*}
Several important observations follow from Fig.~\ref{fig:lbracket_error_comparison}. First, the estimation error generally decreases as the oracle budget increases for all three methods. Second, Bayesian IQAE and MLE-IQAE remain very close across the entire budget range. This confirms that the Bayesian formulation does not degrade the point-estimation accuracy of conventional likelihood-based IQAE. Its main additional benefit is the posterior uncertainty information discussed later.

The most important trend is the increasing advantage of IQAE over direct Monte Carlo sampling as the event becomes rarer. For the case $p_f=1.001\times10^{-2}$, Monte Carlo already observes failures frequently enough to provide a reasonable estimate, although the IQAE estimators still achieve smaller error. As the failure probability decreases to $p_f=1.038\times10^{-3}$ and $p_f=1.221\times10^{-4}$, the limitation of direct sampling becomes more apparent. In the rarest case, Monte Carlo still has a mean relative error of approximately $48.7\%$ at the largest budget, while Bayesian IQAE reduces this error to approximately $8.8\%$.

Table~\ref{tab:lbracket_sampling_statistics} provides additional insight
into the behavior observed in
Figure~\ref{fig:lbracket_error_comparison}.
The results demonstrate that the increasing advantage of IQAE-based
estimators is directly related to the scarcity of observed failures in
classical Monte Carlo simulation.
For the largest failure probability,
$p_f=1.001\times10^{-2}$,
Monte Carlo observes a sufficient number of failures even at the smallest
budget, with an average of approximately five failures at
$B=500$ and more than two hundred failures at
$B=2\times10^4$.
Consequently, although Bayesian IQAE remains more accurate,
the performance gap between the methods is relatively modest.
As the failure probability decreases to
$p_f=1.038\times10^{-3}$,
Monte Carlo enters a data-starved regime.
At $B=500$, only 0.4 failures are observed on average and
60\% of the independent runs observe no failures at all.
Even at $B=1000$, half of the Monte Carlo runs contain zero failures.
The resulting Monte Carlo error exceeds 100\%, whereas Bayesian IQAE
reduces the error to approximately 17\%.
The effect becomes even more pronounced for the rarest event,
$p_f=1.221\times10^{-4}$.
At budgets between 500 and 5000, Monte Carlo observes fewer than one
failure on average and the majority of runs contain no failures.
For example, at $B=5000$, the average number of failures is only 0.3,
while 90\% of the runs observe no failures whatsoever.
In this regime, classical Monte Carlo provides little information about
the target probability and consequently exhibits errors exceeding
100\%.
In contrast, Bayesian IQAE continues to reduce the estimation error
systematically as the oracle budget increases, reaching an average error
below 10\% at $B=2\times10^4$.

These results clarify the source of the observed quantum advantage.
The improvement is not merely a consequence of a different estimator,
but rather arises because amplitude amplification transforms an
extremely small failure probability into a measurable success rate.
Even when the original event is too rare to be observed reliably by
Monte Carlo, the amplified measurements remain informative, allowing
Bayesian updating to continue reducing uncertainty and estimation error.


To understand how Bayesian IQAE achieves this reduction, Table~\ref{tab:lbracket_iqae_trace} reports a representative Bayesian IQAE trace for the rare-event case. 
 During the first two iterations, no successes are observed, even after amplification, but the absence of successes still provides information and reduces the admissible interval in $\theta$. As the Grover depth increases, the amplified success probability becomes large enough to measure directly. For example, although the true failure probability is only $1.221\times10^{-4}$, the measured amplified success probability reaches values such as $0.333$, $0.440$, and $0.739$ at later iterations. This is the central advantage of amplitude amplification: an event that is almost invisible under direct sampling becomes measurable after Grover amplification.
The uncertainty measures in Table~\ref{tab:lbracket_iqae_trace} also illustrates the progressive learning behavior of Bayesian IQAE. The width of the admissible interval for the amplitude angle decreases from $8.857\times10^{-2}$ to $2.86\times10^{-3}$, while the normalized $95\%$ credible-interval width decreases from $69.4$ to $0.36$. The former reflects the resolution of branch ambiguities in the IQAE procedure, whereas the latter reflects the continued concentration of posterior probability mass around the inferred failure probability. Similar to the one-dimensional example, the method therefore provides not only a point estimate but also a rigorous and continuously updated characterization of uncertainty.


\subsubsection{Bayesian Posterior Contraction and Uncertainty Quantification}
\label{sec:lbracket_posterior}



\begin{figure*}[t]
\centering
\begin{subfigure}[t]{0.48\textwidth}
\centering
\includegraphics[width=\linewidth]
{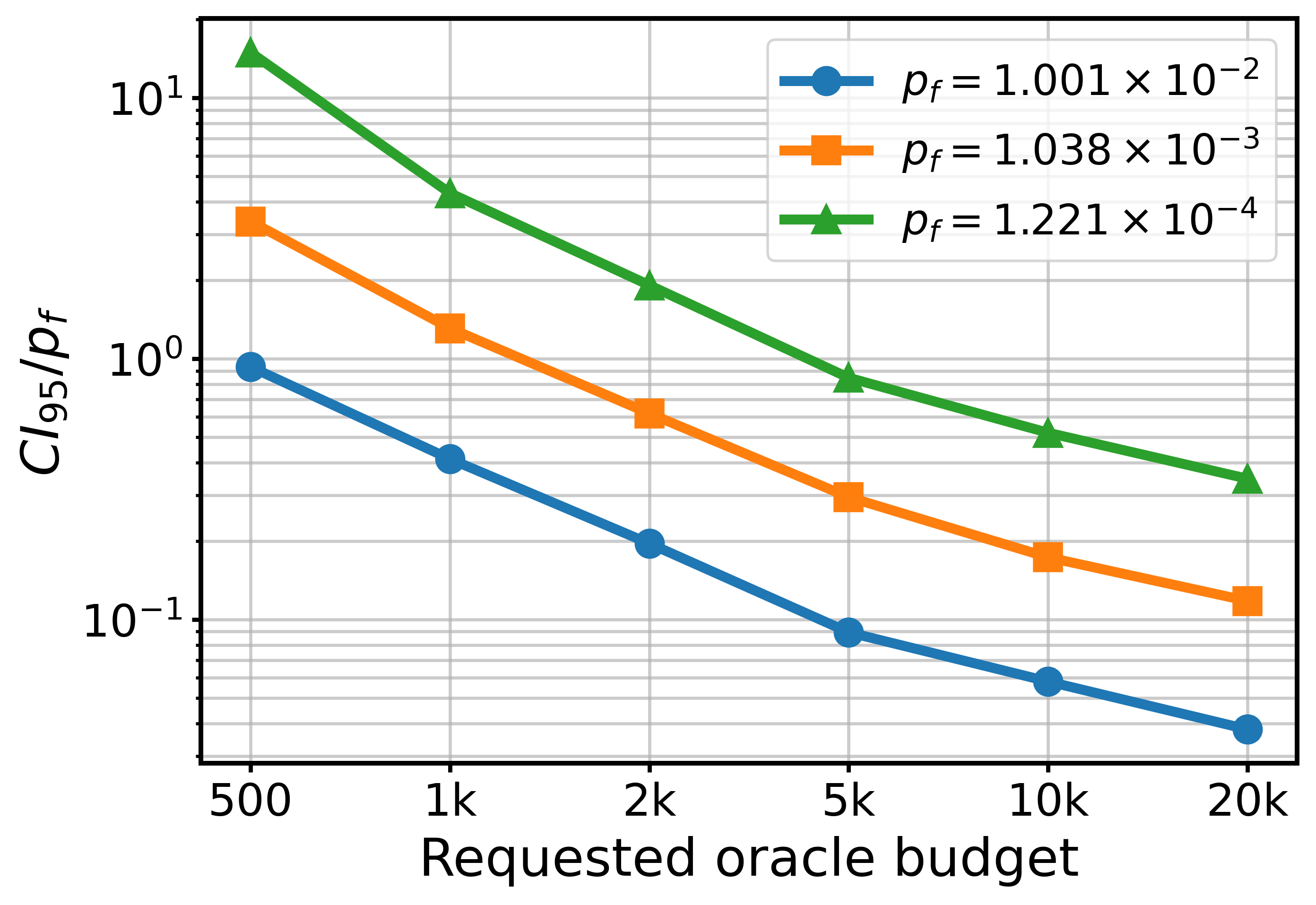}
\caption{ Relative $95\%$ credible-interval width for three failure-probability levels, illustrating systematic posterior contraction as additional oracle evaluations are incorporated. }
\label{fig:relative_credible_width}
\end{subfigure}
\hfill
\begin{subfigure}[t]{0.48\textwidth}
\centering
\includegraphics[width=\linewidth]
{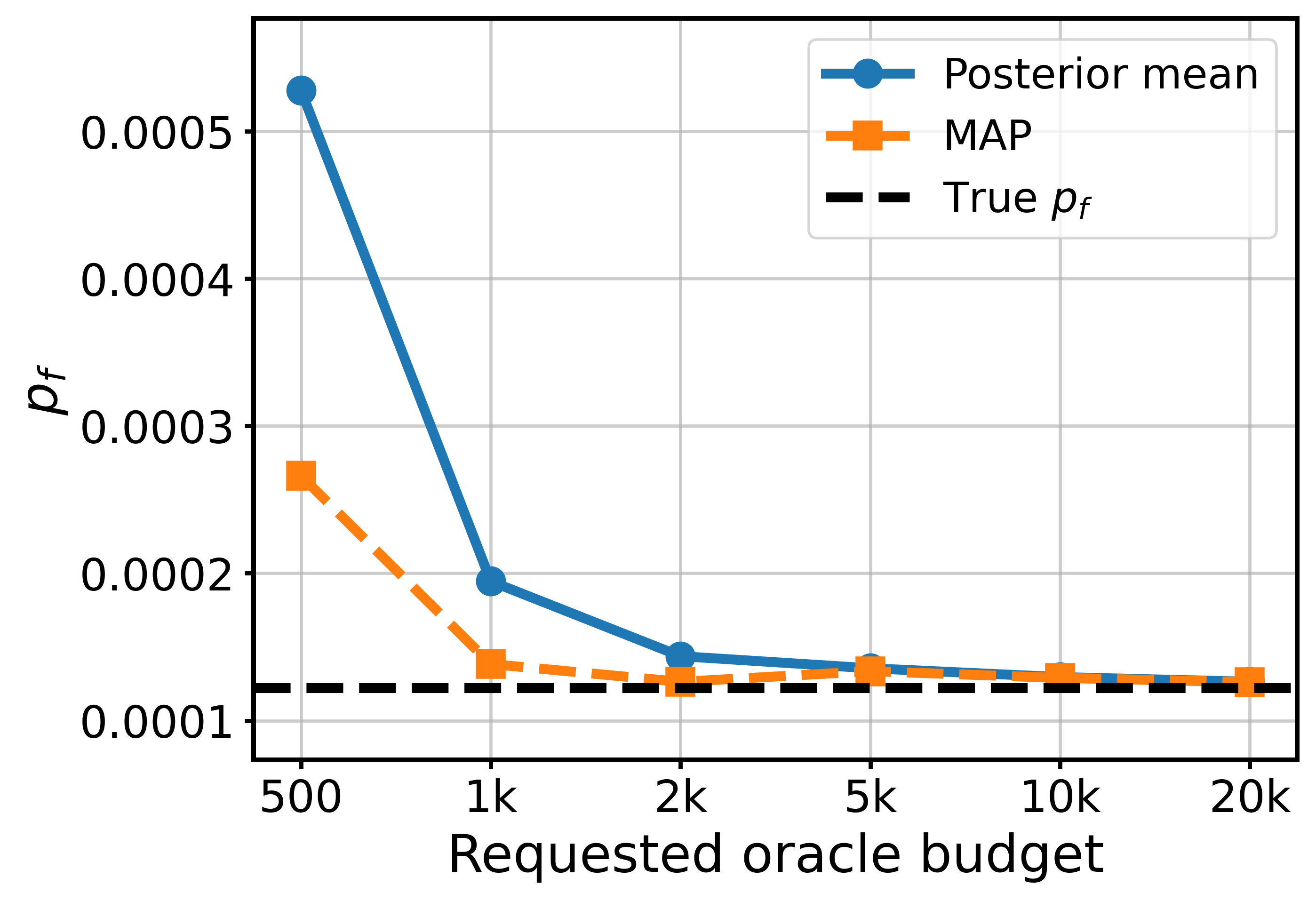}
\caption{Posterior mean and MAP estimate for the rarest failure-probability case, demonstrating convergence toward the reference solution.
}
\label{fig:posterior_mean_ci}
\end{subfigure}
\caption{
Evolution of Bayesian IQAE uncertainty and posterior estimates with increasing oracle budget. 
}
\label{fig:bayesian_uncertainty_convergence}
\end{figure*}

Figure~\ref{fig:relative_credible_width} shows the evolution of the 95\% credible interval as a function of oracle budget for the three failure probabilities considered in this study.
\begin{figure*}[tb]
\centering
\begin{subfigure}[t]{0.32\textwidth}
\centering
\includegraphics[width=\textwidth]
{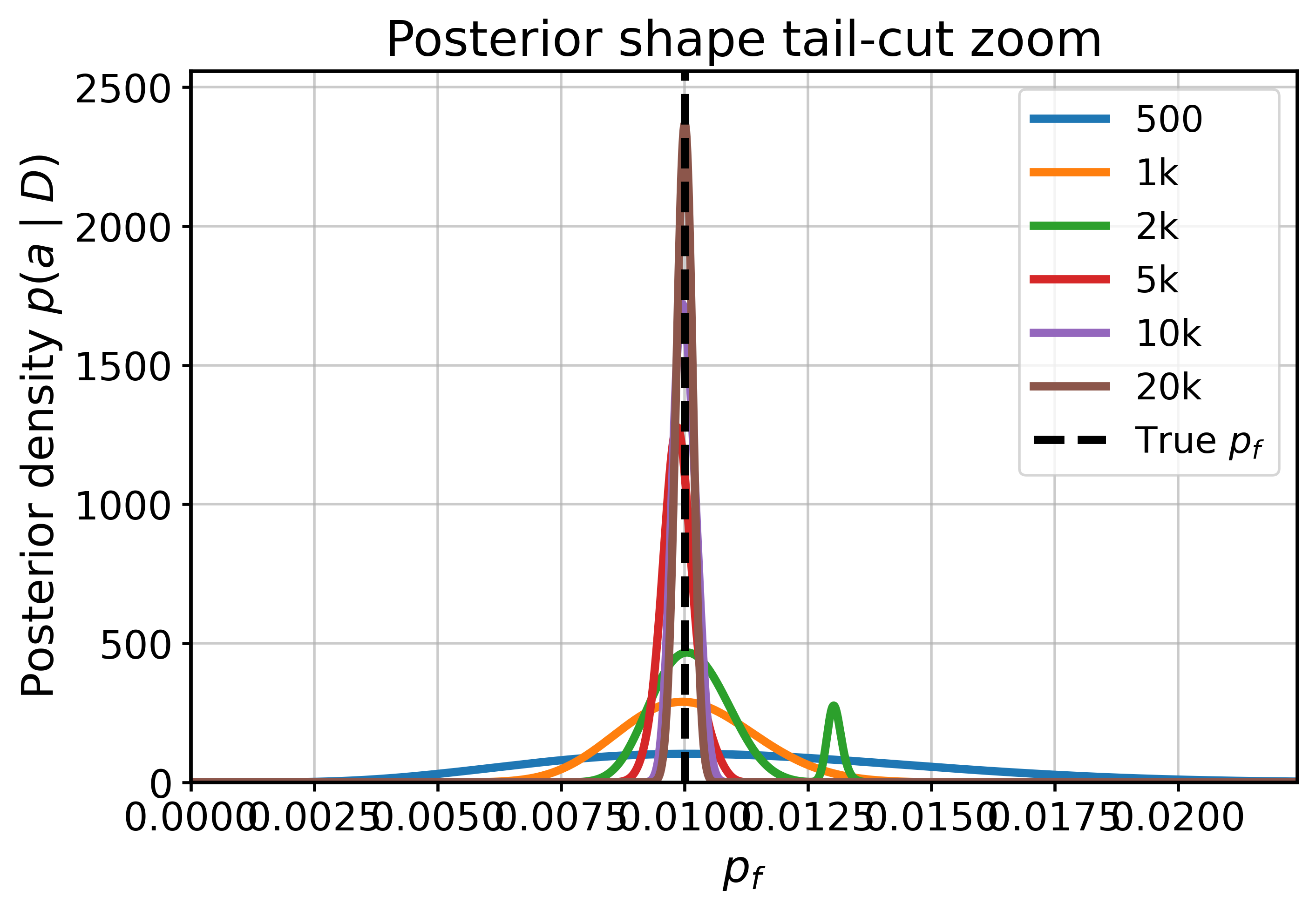}
\caption{$p_f = 1.001\times10^{-2}$}
\end{subfigure}
\hfill
\begin{subfigure}[t]{0.32\textwidth}
\centering
\includegraphics[width=\textwidth]
{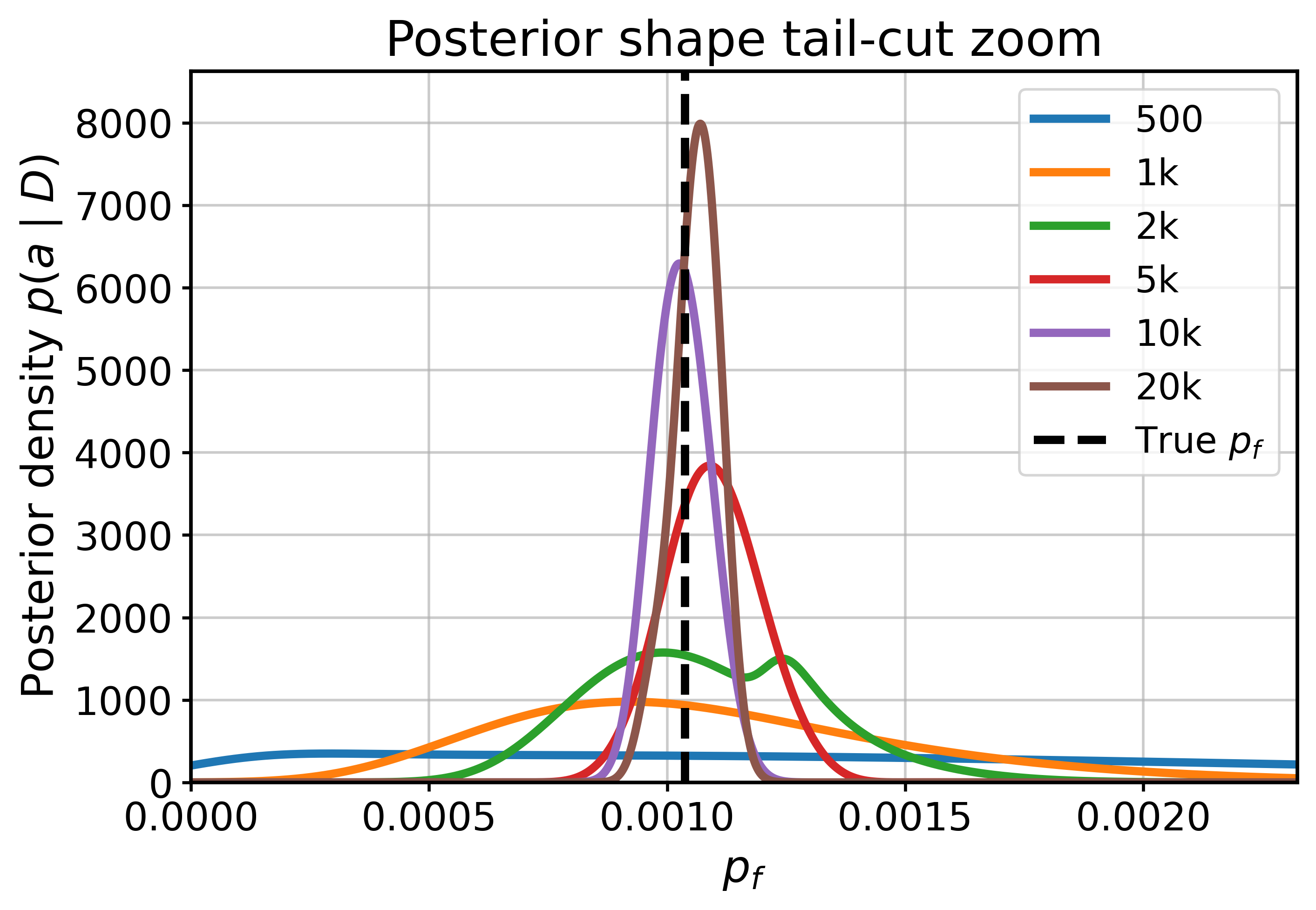}
\caption{$p_f = 1.038\times10^{-3}$}
\end{subfigure}
\hfill
\begin{subfigure}[t]{0.32\textwidth}
\centering
\includegraphics[width=\textwidth]
{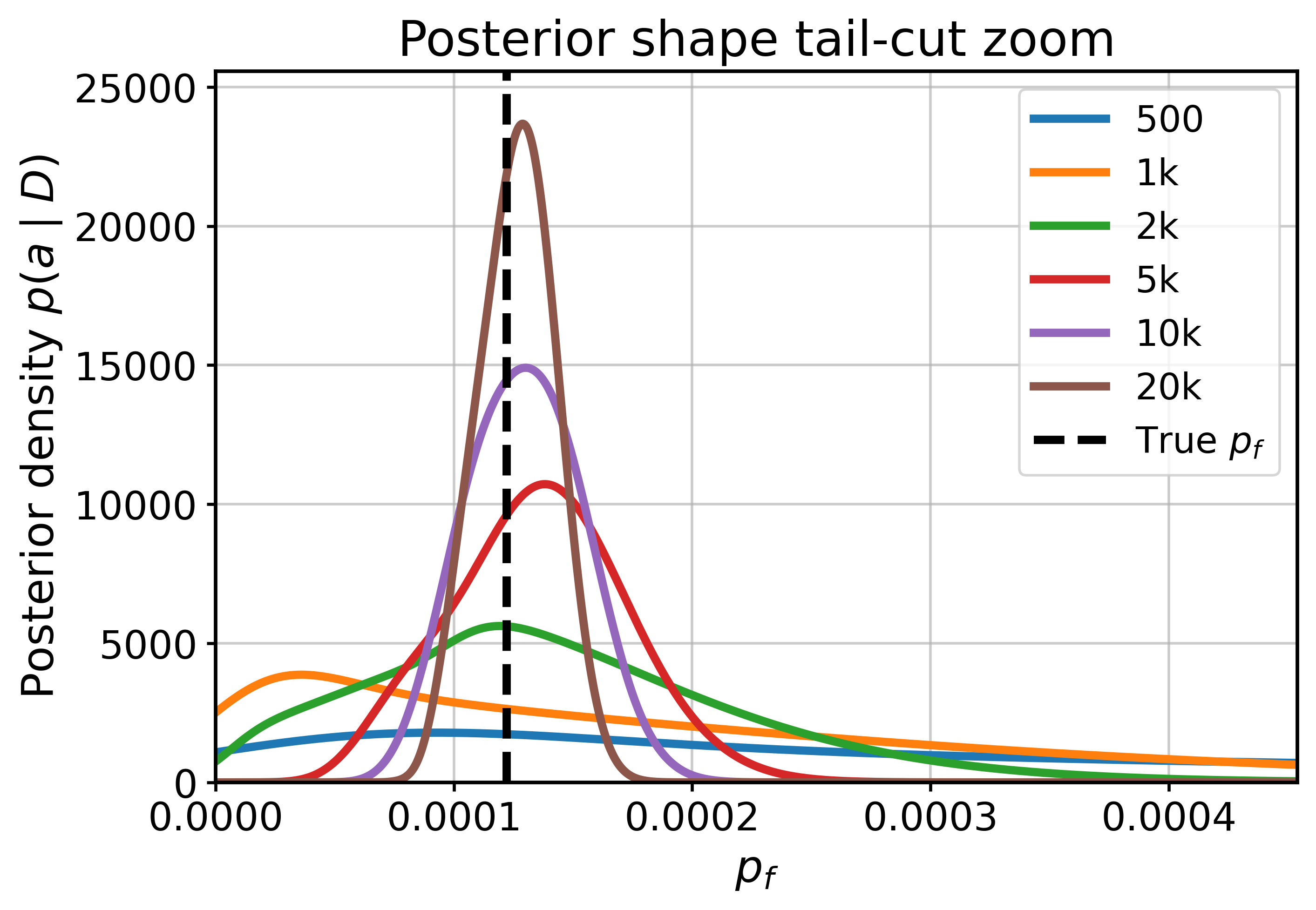}
\caption{$p_f = 1.221\times10^{-4}$}
\end{subfigure}
\caption{
Evolution of Bayesian posterior distributions for three failure probabilities.
As the failure event becomes rarer, the posterior uncertainty increases;
however, systematic posterior contraction is observed for all cases as
the oracle budget increases.
}
\label{fig:posterior_shapes}
\end{figure*}
Several important observations emerge. First, all three curves exhibit monotonic contraction, indicating that Bayesian IQAE continuously accumulates information as additional oracle evaluations become available. Second, the uncertainty level increases systematically as the failure probability decreases. This trend reflects the increasing difficulty of estimating rare events, for which informative observations become progressively less frequent. Nevertheless, even for the rarest event considered, $p_f=1.221\times10^{-4}$, the normalized credible interval width decreases by more than two orders of magnitude over the investigated budget range. The consistent reduction of the relative width of the $95\%$ posterior credible interval, $
\frac{\Delta CI_{95}}
{p_f}$, across all three reliability levels, demonstrates that Bayesian IQAE remains capable of producing increasingly informative posterior distributions even in challenging rare-event regimes.

To examine this behavior in greater detail, Fig.~\ref{fig:posterior_mean_ci} presents the evolution of the posterior mean and maximum-a-posteriori (MAP) estimate for the rarest failure probability.
The posterior mean and MAP estimate converge rapidly toward the reference solution and become nearly indistinguishable at moderate and large oracle budgets. This behavior reflects the increasing concentration and symmetry of the posterior distribution as information accumulates. The close agreement between the posterior mean and MAP estimate also explains the nearly identical performance of Bayesian IQAE and MLE-IQAE observed in Fig.~\ref{fig:lbracket_error_comparison}. As the posterior contracts, the Bayesian estimator naturally approaches the corresponding maximum-likelihood solution. Consequently, the primary advantage of the Bayesian formulation is not improved point-estimation accuracy but rather the ability to characterize uncertainty and quantify confidence in the estimated failure probability.

A more complete picture of the Bayesian learning process is provided by the posterior distributions shown in Fig.~\ref{fig:posterior_shapes}.
This figure provides a direct visualization of Bayesian posterior contraction. For all three reliability levels, the posterior density becomes progressively concentrated around the reference failure probability as the oracle budget increases. At small budgets, the posterior remains relatively broad because only limited information is available regarding the unknown probability. As additional amplified measurements are incorporated, the posterior mass collapses toward a progressively narrower region, indicating increased confidence in the inferred value of $p_f$.

The influence of event rarity is particularly evident when comparing the three panels of Fig.~\ref{fig:posterior_shapes}. The posterior corresponding to $p_f=1.001\times10^{-2}$ becomes sharply concentrated after relatively few oracle evaluations, whereas the rare-event case $p_f=1.221\times10^{-4}$ exhibits substantially broader distributions at comparable budgets. This behavior is expected because smaller failure probabilities generate less informative measurements and therefore require additional amplification and sampling effort. Nevertheless, the contraction observed in all three cases demonstrates that Bayesian IQAE remains robust across a wide range of reliability levels.

\section{Conclusions}
\label{sec:conclusions}

This work developed a Bayesian sequential formulation of iterative quantum amplitude estimation for rare-event structural reliability analysis. The proposed framework reformulates structural failure probability as the expectation of a binary failure indicator and encodes this quantity as a quantum amplitude through a lookup-table oracle constructed from a finite stochastic ensemble. Quantum measurements collected at different Grover amplification depths are assimilated through Bayesian updating over the amplitude angle, producing posterior distributions for the failure probability rather than only point estimates.

The main advantage of the proposed formulation is that it combines the query-efficiency properties of amplitude estimation with uncertainty quantification tools that are directly relevant to reliability analysis. The Bayesian posterior provides posterior means, credible intervals, coefficients of variation, and convergence diagnostics that quantify the remaining uncertainty in the estimated failure probability. This is particularly important in rare-event regimes, where limited measurement budgets may produce sparse or ambiguous observations and where maximum-likelihood estimates alone do not fully characterize estimation confidence.

Numerical results for stochastic finite-element benchmark problems demonstrated that Bayesian IQAE can substantially reduce estimation error relative to direct Monte Carlo simulation under the same idealized oracle-query budget. The results also showed that amplitude amplification converts rare failure events into measurable success probabilities, allowing informative measurements even when the original failure probability is very small. Across the examples considered, the Bayesian estimator achieved point-estimation accuracy comparable to maximum-likelihood IQAE while additionally providing posterior uncertainty quantification and credible intervals. The posterior contraction behavior further showed that the proposed framework provides a transparent mechanism for monitoring convergence as additional oracle evaluations are used.

The present study should be interpreted as a proof of concept for quantum-assisted reliability inference under idealized oracle access. The lookup-table oracle used here isolates the statistical estimation problem from the substantially more difficult task of constructing scalable quantum representations of finite-element models or stochastic structural solvers. Therefore, the reported improvements reflect gains in oracle-query efficiency and statistical inference rather than end-to-end acceleration of structural reliability analysis. Future work should address scalable state preparation, efficient oracle construction for high-dimensional stochastic models, noisy quantum hardware effects, and integration with surrogate or hybrid quantum-classical reliability workflows. Extensions to time-dependent reliability, nonlinear structural response, system reliability, and tail-risk measures such as conditional value-at-risk also provide promising directions for further development.

Overall, the proposed Bayesian IQAE framework provides a statistically interpretable approach for rare-event structural reliability estimation. By producing both failure-probability estimates and quantified uncertainty, it offers a potential pathway toward uncertainty-aware quantum algorithms for risk-informed engineering analysis.

\section{Acknowledgments}
The author acknowledges the financial support of the National Science Foundation under Award No. CMMI-2527378.

\bibliographystyle{elsarticle-num}
\bibliography{bibliography.bib}

\end{document}